\documentclass[
    reprint,
    preprintnumbers,
    superscriptaddress,
    nofootinbib,
     amsmath,amssymb,
     aps,
     prd,
    floatfix,
    ]{revtex4-2}
    
\usepackage{bm}
\usepackage{graphicx}
\usepackage[caption=false]{subfig}
\usepackage[usenames,dvipsnames]{xcolor} 
\usepackage[utf8]{inputenc}
\usepackage{color}
\usepackage[normalem]{ulem} 
\usepackage{physics}
\usepackage{mathrsfs}
\usepackage{comment}
\usepackage[colorlinks]{hyperref}
\hypersetup{ 
	colorlinks=true, 
	linkcolor=black, 
	filecolor=black, 
	citecolor = blue,       
	urlcolor=blue, 
} 
\usepackage{cleveref}
\usepackage{tensor}
\newcommand{\beq}{\begin{equation}}
\newcommand{\eeq}{\end{equation}}

\NewDocumentCommand{\evat}{sO{\bigg}mm}{%
  \IfBooleanTF{#1}
   {\mleft. #3 \mright|_{#4}}
   {#3#2|_{#4}}%
}

\newcommand{\hide}[1]{}

\begin{document}
\count\footins = 1000 

\title{Worldsheet Effective Theory of Abelian-Higgs Cosmic Strings beyond Nambu--Goto}

\author{Sergio Alameda-Calvo}
\email{sergio.alameda@usal.es}
\affiliation{Department of Applied Mathematics, University of Salamanca,
Casas del Parque 2, 37008 -- Salamanca, Spain}
\affiliation{Institute of Fundamental Physics and Mathematics, University of Salamanca, Plaza de la Merced 1, 37008 -- Salamanca, Spain}
\author{Josu C. Aurrekoetxea}	
\email{jaurreko@mit.edu}
\affiliation{Center for Theoretical Physics -- A Leinweber Institute, Massachusetts Institute of Technology, Cambridge, MA 02139, USA}
\author{Jose J. Blanco-Pillado}
\email{josejuan.blanco@ehu.eus}
\affiliation{IKERBASQUE, Basque Foundation for Science, 48011, Bilbao, Spain,}
\affiliation{EHU Quantum Center, University of the Basque Country, UPV/EHU}
\affiliation{Department of Physics, University of the Basque Country, UPV/EHU, 48080, Bilbao, Spain}
\author{\protect\\Alberto Garc\'ia Mart\'in-Caro}\email{alberto.garcia.martin-caro@uvigo.gal}
\affiliation{Instituto de Física, Computación e Ciencias Aeroespaciais (IFCAE), Universidade de Vigo. 32004 Ourense, Spain}
\affiliation{Institute of Theoretical Physics, Jagiellonian University, \L{}ojasiewicza 11, 30-348 Krak\'ow, Poland}
\author{J.M. Queiruga}
\email{xose.queiruga@usal.es}
\affiliation{Department of Applied Mathematics, University of Salamanca,
Casas del Parque 2, 37008 -- Salamanca, Spain}
\affiliation{Institute of Fundamental Physics and Mathematics, University of Salamanca, Plaza de la Merced 1, 37008 -- Salamanca, Spain}


\begin{abstract}

Cosmic strings can support long-lived excitations confined to their cores, 
opening channels for energy exchange with the transverse Goldstone modes 
that describe the string’s motion. We derive a covariant worldsheet theory 
of their coupled dynamics directly from the Abelian-Higgs model at critical 
coupling, with all interaction coefficients fixed by the 
microscopic theory. The resulting action, which we validate against numerical 
lattice field theory simulations, extends the Nambu--Goto description 
to include a massive scalar {\it shape} mode and a massive vector mode. Dualizing 
the vector reveals a geometric distinction: the shape scalar mode couples to 
intrinsic worldsheet curvature, while the pseudoscalar couples to normal bundle 
curvature. These interactions explain the distinct parametric instabilities 
observed in lattice simulations and predict how the relative polarization 
of colliding transverse waves controls the excitation of the two internal 
modes. Our results connect microscopic string structure to the nonlinear 
worldsheet dynamics, providing a predictive framework for cosmic strings 
beyond the Nambu--Goto approximation.
\end{abstract}

\maketitle

\newpage

\section{Introduction}

Cosmic strings are solitonic defects that arise in many extensions of the Standard Model and may have been formed during symmetry breaking phase transitions in the Early Universe \cite{Kibble:1976sj}. Their cosmological signatures probe energy scales inaccessible to terrestrial accelerators and provide a complementary test of fundamental physics \cite{Vilenkin:2000jqa,LIGOScientific:2021nrg,
Planck:2013mgr,NANOGrav:2023hvm,Damour:2000wa,Blanco-Pillado:2017oxo,Blanco-Pillado:2017rnf,Auclair:2019wcv,Blanco-Pillado:2024aca,Lizarraga:2016onn,Caloni:2026dyu,Raidal:2026cpb}. Extracting information about the underlying high energy theory from these signatures requires, however, a controlled connection between the microscopic field theory and the resulting string dynamics.

Cosmic string evolution is commonly described by the Nambu--Goto (NG) action, which governs the massless Goldstone modes associated with broken transverse translations. Spacetime symmetries fix this leading description. A field theory string, however, has finite width and can support massive excitations localized in its core. These modes can be long lived and exchange energy with transverse motion and bulk radiation ~\cite{Blanco-Pillado:2022axf,Alonso-Izquierdo:2024tjc,Aurrekoetxea:2026xrl,Alameda-Calvo:2026dat}. Understanding their influence requires an Effective Field Theory (EFT) that incorporates the internal structure of the string. Such a description should identify how string motion excites the localized modes, how these excitations backreact on the worldsheet, and under which conditions the Nambu--Goto approximation remains reliable.

Our approach is closely analogous to the philosophy of worldline EFT for compact objects where the response of microscopic degrees of freedom is encoded in worldline operators coupled to external fields \cite{Goldberger:2004jt,Porto:2016pyg,Cardoso:2026bfr}. Such a framework provides a systematic description of finite-size effects, resonant excitations, and dissipation within a local EFT on the object's worldline \cite{Goldberger:2005cd,Kobayashi:2025vgl}, and has proven highly successful in describing the response of compact objects to external gravitational fields and the dynamics of compact binaries \cite{Ivanov:2024sds,Bini:2012gu}. 

Here we implement the same idea for string defects, replacing the worldline by a worldsheet and introducing explicit worldsheet fields associated with bound states localized on the string. The lowest order interactions are dictated by symmetries and couple these degrees of freedom to geometric invariants of the induced worldsheet. However, symmetry alone does not determine the localized spectrum or the value of  coupling coefficients.
This requires matching to the underlying four dimensional theory. A microscopic derivation 
is therefore essential for identifying the relevant degrees of freedom and their interactions 
beyond the Nambu--Goto approximation. When localized massive modes are dynamically accessible, 
including them explicitly allows the resulting effective action to describe their production and 
subsequent energy exchange with the translational sector, providing a framework for assessing 
their phenomenological implications.

The Abelian-Higgs (AH) model provides a concrete setting for this program through its Nielsen--Olesen vortex strings ~\cite{Nielsen:1973cs}. Their localized spectrum includes a scalar shape mode ~\cite{Goodband:1995rt,Alonso-Izquierdo:2015tta,Aurrekoetxea:2026xrl} and a distinct vector excitation associated with the gauge field polarization along the string~\cite{Davis:1987df,Arodz:1991ws,Alameda-Calvo:2026dat}. At critical coupling, these modes have equal masses but different dynamical effects. Both lie below the bulk mass threshold, so their decay into bulk radiation proceeds through nonlinear interactions and is suppressed at small amplitudes allowing us to write an effective field theory for them in this regime. Previous work derived the shape mode interactions with the worldsheet geometry, including couplings to its Ricci scalar~\cite{Aurrekoetxea:2026xrl}. The vector mode produces a qualitatively different response in the translational sector ~\cite{Alameda-Calvo:2026dat}, motivating a detailed and unified description that keeps both massive excitations and their couplings to string motion.

In this paper, we derive a covariant worldsheet effective action for the translational, scalar, and vector modes of the Abelian-Higgs string at critical coupling. In order to do that, we express the microscopic fields in coordinates adapted to the string worldsheet, expand in the amplitudes of the internal modes, and integrate over the transverse directions. Matching directly to the microscopic theory determines their mutual interactions and couplings to the worldsheet geometry at the order considered. All coefficients are expressed as transverse integrals involving the vortex background solution and normalized bound state profiles. This construction relates the internal structure of the string to departures from Nambu--Goto evolution and hence provides a framework for assessing the role of localized excitations in cosmic string dynamics.

The effective action considered here describes the conservative dynamics
of the worldsheet degrees of freedom and omits radiation into
bulk fields. Furthermore, its application requires evolution times short
compared with the characteristic radiative damping times of the excited
internal modes. Because their radiative decay arises from
nonlinear interactions, these damping times increase as the excitation
amplitudes decrease, allowing a controlled description over many
oscillation periods in the weak excitation regime. 

The resulting action reveals an important distinction between the two massive bound modes present in the Abelian-Higgs model. The shape scalar couples to the intrinsic Ricci scalar of the worldsheet, while the vector field strength couples to the curvature of the normal bundle, which describes the rotation of the transverse frame along the string. In two spacetime dimensions, the massive vector mode carries one propagating degree of freedom and admits an equivalent pseudoscalar description. In this dual formulation, the two internal excitations couple to the intrinsic and normal curvature densities, respectively. The pseudoscalar interaction has the same geometric structure as the worldsheet axion coupling proposed for confining QCD flux tubes ~\cite{Dubovsky:2013gi}. The Abelian-Higgs vortex thus supplies an explicit microscopic realization of this interaction, with its mass and coupling fixed by the underlying field theory.

We analyze the resulting dynamical consequences and test the effective theory predictions using 
lattice simulations with adaptive mesh refinement. The geometric couplings account for the distinct parametric instabilities previously observed when the scalar and vector modes are excited ~\cite{Aurrekoetxea:2026xrl,Alameda-Calvo:2026dat}. Their expansion about a straight string reproduces the corresponding collective coordinate interactions and explains the different polarization patterns of the induced transverse motion.

Collisions of counterpropagating transverse waves provide a complementary test in which the incoming string carries no internal excitation. Nambu--Goto dynamics determines the leading motion and scattering time delay, while the curvature generated during the collision sources the massive modes. The intrinsic and normal curvature couplings predict specific dependences on the relative response of the shape and vector mode on the polarization of the incoming waves. Comparing these predictions with the internal amplitudes extracted from the lattice simulations tests how worldsheet geometry transfers energy into the microscopic degrees of freedom of the string.

A direct interaction between the two internal modes also predicts the shape excitation generated by a coherent vector oscillation. We isolate this process numerically and compare its amplitude and time dependence with the effective action description.


The outline of the paper is the following. We begin by reviewing the vortex background and its localized excitation spectrum. We then derive the covariant worldsheet action from the microscopic theory and dualize the vector mode into a pseudoscalar. Next, we compare the predictions of the effective theory with lattice simulations in three settings: the parametric excitation of the Goldstone modes by a homogeneous bound mode, the excitation of the bound modes by colliding transverse waves, and the direct coupling between the vector and shape modes. We conclude by summarizing our results and discussing their implications for string dynamics. The appendices contain the technical details: Appendix \ref{app-Effective-Action-Computation} provides the detailed microscopic derivation of the effective action, Appendix \ref{app-CC-vs-EFT} compares it with the collective coordinate approach, Appendix \ref{app-colliding-wiggles} provides an in depth discussion on the Nambu--Goto scattering of the Goldstone modes
analytically, and derives the bound mode response it induces. Finally, Appendix \ref{app-convergence-test} describes the numerical methodology and convergence tests.

\section{String background and excitation spectrum} 
In this paper, we consider the Abelian-Higgs model described by the $4d$ action 
\begin{equation}
\label{eq:FT-action}
\begin{aligned}
S=-\int d^4x\,\biggl[&
\frac14 F_{\mu\nu}F^{\mu\nu}
+\frac12(D_\mu\phi)^*D^\mu\phi
+\frac{\lambda}{8}\bigl(|\phi|^2-\eta^2\bigr)^2
\biggr],
\end{aligned}
\end{equation}
where $D_\mu=\partial_\mu-ieA_\mu$,
$F_{\mu\nu}=\partial_\mu A_\nu-\partial_\nu A_\mu$, and we will use
$\eta_{\mu\nu}=\operatorname{diag}(-,+,+,+)$.
The symmetry breaking vacuum $|\phi|=\eta$ has Higgs and gauge masses given
by $m_H=\sqrt{\lambda}\,\eta$ and $m_A=e\eta$. For simplicity, 
we introduce the dimensionless quantities $\tilde{x}^\mu=e\eta x^\mu$,
$\tilde{\phi}=\phi/\eta$, and $\tilde{A}_\mu=A_\mu/\eta$, 
and we will suppress tildes below. Thus  the action carries an overall
factor $e^{-2}$ and the classical equations depend on 
$\beta=\lambda/e^2 = (m_H/m_A)^2$ where masses
and frequencies are measured in units of $e\eta$. One can show that within this 
theory, a static, straight, unit winding Nielsen--Olesen string along the $z$ direction
can be described by the following ansatz
\begin{equation}
\label{eq:ah-background}
\bar\phi=f(\rho)e^{i\theta},\qquad
\bar A_i=\alpha(\rho)\partial_i\theta,\qquad
\bar A_t=\bar A_z=0,
\end{equation}
where $i=x,y$, $\rho^2=x^2+y^2$, are the transverse directions to the string, 
and $\theta$ is the polar angle~\cite{Nielsen:1973cs}.
Regularity and finite tension require $f(0)=\alpha(0)=0$ and
$f(\infty)=\alpha(\infty)=1$.
At critical coupling, $\beta=1$, the string saturates the
Bogomolny-Prasad-Sommerfield (BPS) bound and the profile functions obey
\begin{equation}
\label{eq:ah-bps-profiles}
f'=\frac{1-\alpha}{\rho}f,\qquad
\alpha'=\frac{\rho}{2}(1-f^2).
\end{equation}
where primes denote derivatives with respect to the radial coordinate perpendicular
to the string, $\rho$.

Physical linear perturbations around the string solution and proportional to
$e^{-i\Omega t+ikz}$ satisfy the dispersion relation
$\Omega^2=k^2+\omega_{\rm I}^2$, where $\omega_{I}$
denotes the worldsheet rest mass for the three different kinds of normal modes present in the 
Abelian-Higgs model string. The two translational \emph{zero modes} (Goldstone modes) correspond to displacements of the vortex in the
transverse plane. They have $\omega_0=0$ and propagate
as massless transverse waves. In addition, the BPS string
in the Abelian-Higgs model supports only two other localized
massive modes.

The scalar \emph{shape mode} that describes an axially
symmetric deformation of the
form~\cite{Goodband:1995rt,Kojo:2007bk,Alonso-Izquierdo:2015tta,Aurrekoetxea:2026xrl}
\begin{equation}
\label{eq:ah-shape-mode}
\delta\phi=\chi(t,z)\varphi(\rho)e^{i\theta},\qquad
\delta A_i=\chi(t,z)\rho a(\rho)\partial_i\theta,
\end{equation}
with $\delta A_t=\delta A_z=0$. Here, the real functions
$\varphi(\rho)$ and $a(\rho)$ specify its transverse
profiles, while the real worldsheet massive scalar $\chi(t,z)$
describes its amplitude and propagation along the string. 

The \emph{vector mode} has components along the
worldsheet coordinates $\sigma^A=(t,z)$,
\begin{equation}
\label{eq:ah-vector-mode}
\delta A_A=C_A(t,z)\psi(\rho),
\end{equation}
with $\delta\phi=\delta A_i=0$. Here $\psi(\rho)$ is
the radial profile, and the worldsheet vector $C_A(t,z)$
satisfies the constraint $\partial^A C_A=0$.

This mode carries one physical massive vector polarization whose worldsheet vector mode
is described by $C_A$ and its radial profile satisfies~\cite{Alameda-Calvo:2026dat}
\begin{equation}
\label{eq:ah-vector-spectrum}
-\psi''-\frac{1}{\rho}\psi'+f^2\psi=\omega_v^2\psi~.
\end{equation}
The bound solution is regular and square integrable so we can normalize it by choosing
\begin{equation}
\label{eq:normalization-psi}
2\pi\int_0^\infty \rho\,d\rho\,\psi^2=1~.
\end{equation}
At critical coupling, factorization of the fluctuation operators
relates the shape and vector mode profiles and establishes their
mass degeneracy~\cite{Alameda-Calvo:2026dat}:
\begin{equation}
\label{eq:ah-bps-mode-pair}
\varphi=\frac{f\psi}{\omega_b},\qquad
a=-\frac{\psi'}{\omega_b},\qquad
\omega_s=\omega_v\equiv\omega_b.
\end{equation}
Here $\omega_s$ and $\omega_v$ denote the worldsheet rest masses
of the shape and vector modes, respectively, and $\omega_b$
denotes their common value throughout the remainder of this work.
The mode relations in Eq.~\eqref{eq:ah-bps-mode-pair}, together
with the eigenvalue equation~\eqref{eq:ah-vector-spectrum},
imply, after integration by parts,
\begin{equation}
2\pi\int_0^\infty \rho\,d\rho\,(\varphi^2+a^2)
=2\pi\int_0^\infty \rho\,d\rho\,\psi^2=1,
\end{equation}
where the boundary terms vanish by regularity and normalizability,
and the final equality follows from
Eq.~\eqref{eq:normalization-psi}.
Numerically, $\omega_b\simeq0.882$~\cite{Alameda-Calvo:2026dat}.
Restoring dimensions, the localized modes have masses
$m_0=0$ and $m_s=m_v\simeq0.882\,e\eta$.
Both massive modes lie below the bulk threshold
$m_H=m_A=e\eta$ and therefore cannot radiate into the bulk
at linear order. Their classical decay is mediated by nonlinear
interactions and is suppressed at small amplitudes, allowing
these excitations to persist for many oscillation 
periods~\cite{Alonso-Izquierdo:2015tta,Alameda-Calvo:2026dat}.

\section{Microscopic derivation of the worldsheet effective action}

To separate the motion of the string from changes in its internal
structure, we introduce coordinates adapted to the string worldsheet,
following the method introduced in Ref.~\cite{Blanco-Pillado:2024bev,Aurrekoetxea:2026xrl}. The embedding
$X^\mu(\sigma^A)$, with $A=\lbrace{0,1\rbrace}$, specifies the string position and
defines a local decomposition into tangent and normal directions.
In the local rest frame of the string, we approximate the fields
in each normal plane by the straight BPS vortex profiles supplemented
by the retained bound modes. Substituting this ansatz into the
microscopic four dimensional action and integrating over the
transverse coordinates yields the effective worldsheet action
and its geometric couplings. We outline the construction of this effective action here and present the detailed derivation
of each term in Appendix~\ref{app-Effective-Action-Computation}.

Let us start by introducing some geometric quantities related to the
string worldsheet, in particular the tangent vectors $e_A^\mu=\partial_A X^\mu$ and the oriented spacelike normals satisfying
$n_i^\mu n_{j\mu}=\delta_{ij}$ and $e_A^\mu n_{i\mu}=0$,
where Greek indices refer to all four adapted
coordinates, $A,B=\lbrace{0,1\rbrace}$ to the worldsheet, and $i,j=\lbrace{1,2\rbrace}$ to the
normal directions.

Using these vectors we can 
describe the position of points near the string by
\begin{equation}
\label{eq:eft-adapted-coordinates}
y^\mu(\sigma,u)
=
X^\mu(\sigma)+u^i n_i^\mu(\sigma),
\qquad
\rho^2=\delta_{ij}u^iu^j.
\end{equation}
This construction applies when the curvature radii are large
compared with the transverse widths of the specified modes.
The induced metric is $\gamma_{AB}=e_A^\mu e_{B\mu}$.
Bending of the worldsheet and rotation of its normal frame (the so-called 
{\it twist}) are described by
$K^i_{AB}=-n^i_\mu\partial_A e_B^\mu$ and
$\omega_A=n_{1\mu}\partial_A n_2^\mu$, respectively.
Writing $K^i=\gamma^{AB}K^i_{AB}$, we define the intrinsic
Ricci scalar and the curvature of the normal connection by
\begin{equation}
\label{eq:eft-geometric-invariants}
\mathcal R=K^iK_i-K^i_{AB}K_i^{AB},
\qquad
\Omega_{AB}=\nabla_A\omega_B-\nabla_B\omega_A.
\end{equation}

To compute the effective action, we allow the amplitudes of the
shape and vector modes to vary along the string. Choosing the
Higgs phase to be the polar angle $\theta(u)$, we write the
fields in adapted coordinates as
\begin{equation}
\label{eq:eft-mode-lift}
\begin{aligned}
\phi&=[f(\rho)+\varphi(\rho)\chi(\sigma)]e^{i\theta},
\\
A_i&=[\alpha(\rho)+\rho a(\rho)\chi(\sigma)]\partial_i\theta,
\\
A_A&=[1-\alpha(\rho)-\rho a(\rho)\chi(\sigma)]\omega_A
+\psi(\rho)C_A(\sigma),
\end{aligned}
\end{equation}
where the radial profiles keep the normalizations of the
preceding section, while $\chi$ and $C_A$ describe the internal
worldsheet excitations. Even in the absence of these excitations, the
background contains the component
$\bar A_A=(1-\alpha)\omega_A$.
This component accounts for the rotation of the normal frame
and ensures covariance under the accompanying $U(1)$ gauge
transformation.

Inserting Eq.~\eqref{eq:eft-mode-lift} into the microscopic
four dimensional action and integrating over the transverse directions, 
$u^i$, determines the effective couplings
\cite{Aurrekoetxea:2026xrl,Alameda-Calvo:2026dat}.
We organize the calculation of the effective action through 
successive field variations at a fixed embedding. The contribution 
of degree $n$ in $\chi$ and $C_A$ is $S^{(n)}=\delta^nS/n!$, evaluated on the background vortex configuration. The successive variations and 
their reduction are presented in Appendix~\ref{app-Effective-Action-Computation}.

At zeroth order in the internal amplitudes, the background
energy gives the Nambu--Goto action with dimensionless tension
$\mu=\pi$, corresponding to physical tension $\pi\eta^2$.
It also generates an Euler term with a constant coefficient,
which does not affect the local equations of motion and is
only included below for completeness.

The first variation need not vanish because the vortex ansatz
is not an exact solution for a general curved worldsheet.
At linear order, the straight vortex equations cancel the
terms independent of curvature.
The remaining terms describe how the string geometry excites
its internal modes. The transverse measure produces the scalar
coupling $\chi\mathcal R$, while the cross term between
$\bar F_{AB}=(1-\alpha)\Omega_{AB}$ and the vector fluctuation
$\delta F_{AB}=\psi G_{AB}$ produces a term of the form
$G_{AB}\Omega^{AB}$.
Here
$G_{AB}=\nabla_A C_B-\nabla_B C_A$
is the worldsheet field strength of the vector mode.
Note that this coupling respects the local rotation symmetry of the
normal frame, since the twist enters through the invariant
normal bundle curvature $\Omega_{AB}$.

At quadratic order, the transverse eigenvalue equations
combine the radial gradient and potential contributions
into worldsheet mass terms.
This gives a massive scalar and a Proca field, with kinetic
coefficients fixed by the profile normalizations and a common
mass $\omega_b$ fixed by the transverse eigenvalue problem.

At cubic order, a direct interaction arises from the
scalar kinetic term. The shape deformation
changes the Higgs profile and hence the mass term experienced
by the vector mode, producing an interaction proportional to
$\chi C_A C^A$. At the same order, the scalar kinetic term
and the Higgs potential generate a scalar self interaction
proportional to $\chi^3$, which we also include for consistency
of the cubic truncation. Note that the BPS profile equations
relate the cubic scalar self interaction to the mixed
scalar-vector coupling $g_{SV}$.

Collecting the leading quadratic terms, the geometric sources,
and the cubic interaction gives
\begin{align}
S_{\rm eff}
=-\frac{1}{e^2}\!\int \!d^2\sigma\,\sqrt{-\gamma}\,
\!\biggl[\notag
&\mu+\nu\mathcal R
-\kappa_\Omega\Omega_{AB}\Omega^{AB}\notag
\\
&+\frac12\nabla_A\chi\nabla^A\chi
+\frac{\omega_b^2}{2}\chi^2 \label{eq:eft-combined-bps-action}
\\
&+\frac14G_{AB}G^{AB}
+\frac{\omega_b^2}{2}C_A C^A
\notag
\\
&-\kappa_S\chi\mathcal R
+\frac{\kappa_V}{2}G_{AB}\Omega^{AB}\notag
\\
&+g_{SV}\chi C_A C^A + \frac{1}{2}g_{SV} \chi^3
\biggr]+\cdots,\notag
\end{align}
where the coefficients are determined by the background
and bound state profiles,
\begin{equation}
\label{eq:eft-matched-coefficients}
\begin{aligned}
\mu
&=2\pi\int_0^\infty d\rho\,
\left(\rho f'^2+\frac{\alpha'^2}{\rho}\right)
=\pi,
\\
\nu
&=\frac{\pi}{2}\int_0^\infty d\rho\,
\left(\rho^3f'^2+\rho\alpha'^2\right)
=\frac{\pi}{2},
\\
\kappa_S
&=\pi\int_0^\infty d\rho\,
\left(\rho^2f'\varphi+\rho\alpha'a\right)
\approx1.2005,
\\
\kappa_V
&=2\pi\int_0^\infty \rho\,d\rho\,
(1-\alpha)\psi
\approx2.7228,
\\
\kappa_\Omega
&=-\frac{\pi}{2}\int_0^\infty \rho\,d\rho\,
(1-\alpha)^2
\approx-1.9503,
\\
g_{SV}
&=2\pi\int_0^\infty \rho\,d\rho\,
f\varphi\psi^2
\approx0.0864.
\end{aligned}
\end{equation}

The ellipsis in Eq. \eqref{eq:eft-combined-bps-action} includes additional 
geometric corrections and
interactions among the internal modes.
In a simultaneous weak field expansion about a straight
string, the transverse displacement, $\chi$, and $C_A$
are all counted as first order, whereas $\mathcal R$ and
$\Omega_{AB}$ begin at second order.
The displayed geometric couplings and the direct mixed
interaction therefore enter at cubic order in this counting.
The detailed derivation of this effective action, including the additional amplitude
and curvature terms, is given in
Appendix~\ref{app-Effective-Action-Computation}.

\section{Dualization of the vector mode}
\label{sec:dualization}

In two spacetime dimensions, a massive vector field carries
a single propagating degree of freedom and admits an equivalent
pseudoscalar description. We derive this dual formulation and 
describe its geometric interpretation, keeping the dependence on 
the shape mode amplitude explicit.

We begin with the terms in the effective action involving the 
worldsheet vector,
\begin{align}
\notag
S_V
=-\frac{1}{e^2}\int d^2\sigma\,\sqrt{-\gamma}\,
\biggl[
&\frac14G_{AB}G^{AB}
+\frac12M^2(\chi)C_A C^A
\\
&+\frac{\kappa_V}{2}G_{AB}\Omega^{AB}
\biggr],
\label{eq:dual-vector-sector}
\end{align}
where
\begin{equation}
\label{eq:dual-effective-vector-mass}
M^2(\chi)=\omega_b^2+2g_{SV}\chi
\end{equation}
includes the mixed interaction between the shape and vector
modes, and as we mentioned earlier, 
${G_{AB}=\nabla_A C_B-\nabla_B C_A}$.

Now, in order to dualize this theory, we treat $G_{AB}$ as an 
independent antisymmetric tensor and introduce a pseudoscalar Lagrange multiplier $\vartheta$
through the term
\begin{equation}
\label{eq:dual-first-order-constraint}
\begin{aligned}
\Delta S\!
=\!\frac{\omega_b}{2e^2}
\!\int\! d^2\sigma\,\sqrt{-\gamma}\,
\vartheta\varepsilon^{AB}
\left(G_{AB}\!-\!\nabla_A C_B\!+\!\nabla_B C_A\right).
\end{aligned}
\end{equation}
Varying with respect to $\vartheta$ enforces the original
relation between $G_{AB}$ and $C_A$, since an antisymmetric
tensor in two dimensions has only one independent component.

Using the antisymmetry of $\varepsilon^{AB}$ and integrating
the terms containing derivatives of $C_A$ by parts gives,
up to a boundary term, 
\begin{align}
S_V \!+ \Delta S
=\!-\frac{1}{e^2}\!\!\int \!\!d^2\sigma\,\sqrt{-\gamma}\,
\biggl[
&\frac14G_{AB}G^{AB}
+\frac12M^2(\chi)C_A C^A
\notag\\
&+\frac{\kappa_V}{2}G_{AB}\Omega^{AB}
\notag\\
&-\frac{\omega_b}{2}
\vartheta\varepsilon^{AB}G_{AB}
\label{eq:dual-first-order-action}
\\
&-\omega_b\varepsilon^{AB}
(\nabla_A\vartheta)C_B
\biggr].\notag
\end{align}
Both $G_{AB}$ and $C_A$ now appear without derivatives.
Their algebraic equations give
\begin{equation}
\label{eq:dual-field-relations}
\begin{aligned}
G_{AB}
&=\omega_b\vartheta\varepsilon_{AB}
-\kappa_V\Omega_{AB},
\\
C^B
&=\frac{\omega_b}{M^2(\chi)}
\varepsilon^{AB}\nabla_A\vartheta.
\end{aligned}
\end{equation}
Contracting the first relation with $\varepsilon^{AB}$ yields
\begin{equation}
\label{eq:dual-pseudoscalar-identification}
\vartheta
=
-\frac{1}{2\omega_b}
\varepsilon^{AB}
\left(G_{AB}+\kappa_V\Omega_{AB}\right).
\end{equation}
The contraction with the Levi-Civita tensor identifies
$\vartheta$ as a pseudoscalar. We define the dual normal curvature by
\begin{equation}
\label{eq:dual-normal-curvature}
\mathcal T
\equiv\frac12\varepsilon^{AB}\Omega_{AB}
=
\frac12\varepsilon^{AB}\epsilon_{ij}
K^i_{AC}K^{jC}_{B},
\end{equation}
where the second equality follows
from the Ricci equation in flat spacetime.
On the algebraic solutions
\eqref{eq:dual-field-relations}, the terms involving
$G_{AB}$ in the bracket of
Eq.~\eqref{eq:dual-first-order-action} combine as
\begin{equation}
\label{eq:dual-field-strength-reduction}
\begin{aligned}
&\frac14G_{AB}G^{AB}
+\frac{\kappa_V}{2}G_{AB}\Omega^{AB}
-\frac{\omega_b}{2}\vartheta\varepsilon^{AB}G_{AB}
\\
&\qquad
=-\frac14G_{AB}G^{AB}
\\
&\qquad
=\frac{\omega_b^2}{2}\vartheta^2
+\omega_b\kappa_V\vartheta\mathcal T
-\frac{\kappa_V^2}{4}\Omega_{AB}\Omega^{AB}.
\end{aligned}
\end{equation}
Similarly, the terms involving $C_A$ give
\begin{equation}
\label{eq:dual-vector-reduction}
\begin{aligned}
&\frac12M^2(\chi)C_A C^A
-\omega_b\varepsilon^{AB}(\nabla_A\vartheta)C_B
\\
&\qquad
=-\frac12M^2(\chi)C_A C^A
\\
&\qquad
=\frac{\omega_b^2}{2M^2(\chi)}
\nabla_A\vartheta\nabla^A\vartheta,
\end{aligned}
\end{equation}
where again we have used
Eq.~\eqref{eq:dual-field-relations}.

Substituting these expressions into the first order action
gives the dual vector sector,
\begin{equation}
\label{eq:dual-vector-action}
\begin{aligned}
S_{{\rm dual},V}
=-\frac{1}{e^2}\int d^2\sigma\,\sqrt{-\gamma}\,
\biggl[
&\frac{\omega_b^2}{2M^2(\chi)}
\nabla_A\vartheta\nabla^A\vartheta
\\
&+\frac{\omega_b^2}{2}\vartheta^2
+\omega_b\kappa_V\vartheta\mathcal T
\\
&-\frac{\kappa_V^2}{4}
\Omega_{AB}\Omega^{AB}
\biggr].
\end{aligned}
\end{equation}
This expression is exact for the vector terms displayed
in Eq.~\eqref{eq:dual-vector-sector}.
We restrict ourselves to the regime where $M^2(\chi)>0$, for which the pseudoscalar
kinetic coefficient is positive.
About the unexcited straight string, where $\chi=0$,
the chosen normalization gives $\vartheta$ the same
quadratic kinetic normalization as $\chi$ and mass
$\omega_b$.

To recover the expansion through cubic order in the
worldsheet fields, we assume
$|2g_{SV}\chi|\ll\omega_b^2$ and expand
\begin{equation}
\label{eq:dual-kinetic-expansion}
\frac{\omega_b^2}{2M^2(\chi)}
=
\frac12-\frac{g_{SV}}{\omega_b^2}\chi
+\mathcal O(\chi^2).
\end{equation}
This expansion is consistent with the small amplitude
approximation used in constructing the effective action.
The term linear in $\chi$ generates the mixed derivative
interaction
$-g_{SV}\chi\nabla_A\vartheta\nabla^A\vartheta/\omega_b^2$
inside the action integrand.

Restoring the remaining terms in
Eq.~\eqref{eq:eft-combined-bps-action}, including the
cubic scalar self-interaction $\frac{1}{2}g_{SV}\chi^3$, gives

\begin{align}
S_{\rm dual}\!
=\!-\frac{1}{e^2}\!\!\int\!\! d^2\sigma\sqrt{-\gamma}\,
\!\biggl[\notag
&\mu+\nu\mathcal R
\\
&+\frac12\nabla_A\chi\nabla^A\chi
+\frac12\nabla_A\vartheta\nabla^A\vartheta\notag
\\
&+\frac{\omega_b^2}{2}
\left(\chi^2+\vartheta^2\right)
+\frac{1}{2}g_{SV}\chi^3\notag
\\
&-\kappa_S\chi\mathcal R
+\omega_b\kappa_V\vartheta\mathcal T\notag
\\
&-\frac{g_{SV}}{\omega_b^2}
\chi\nabla_A\vartheta\nabla^A\vartheta
\label{eq:dual-combined-bps-action}
\\
&-\left(\kappa_\Omega+\frac{\kappa_V^2}{4}\right)
\Omega_{AB}\Omega^{AB}
\biggr]+\cdots.\notag
\end{align}

The contribution proportional to $\kappa_V^2/4$ is
the local contact term generated by eliminating
$G_{AB}$.
It shifts the coefficient of the
$\Omega_{AB}\Omega^{AB}$ term already present in
Eq.~\eqref{eq:eft-combined-bps-action}, whose
$\kappa_\Omega$ contribution is inherited from the
microscopic reduction. The ellipsis denotes additional 
contributions at this geometric order and interactions 
beyond the order considered in the amplitude expansion.

\subsection{Geometric interpretation of the couplings}

The dual formulation identifies the geometric origin
of the two sources. The scalar $\chi$ couples to the
intrinsic curvature $\mathcal R$, while the pseudoscalar
$\vartheta$ couples to the normal curvature
$\mathcal T=\tfrac12\varepsilon^{AB}\Omega_{AB}$.
These quantities represent, up to normalization, the
Euler densities of the tangent and normal bundles,
respectively. On a closed, oriented Euclidean worldsheet,
their integrals yield the Euler characteristic and the
normal Euler number, the latter being related to the
signed self-intersection number~\cite{Polyakov:1986cs}.
With constant coefficients, both contributions are
topological and leave the local classical equations
of motion unchanged, although they may affect boundary
terms and the relative weighting of topological sectors.
Their coupling to dynamical fields generates local
interactions that mediate energy exchange between
the string geometry and its internal excitations.

At critical coupling, the geometric coefficients satisfy
$\omega_b\kappa_V=2\kappa_S$, as derived in
Sec.~\ref{sub-sec:kappas-relation} of
Appendix~\ref{app-Effective-Action-Computation}.
With the field and curvature normalizations adopted here,
the magnitude of the pseudoscalar coupling is therefore
twice that of the scalar coupling\footnote{Note that in terms 
of the equally normalized curvature densities $\mathcal R/2$ 
and $\mathcal T$, the two internal modes couple with the same 
strength.}. Together with the
mass degeneracy, this relation suggests a common origin
in the BPS structure of the microscopic vortex. 

The pseudoscalar interaction has the same geometric
form as the coupling introduced for the
\emph{worldsheet axion} of confining QCD flux
tubes~\cite{Dubovsky:2013gi,Dubovsky:2015zey}.
In that context, a massive pseudoscalar coupled to
the transverse Goldstone modes accounts for features
of the lattice excitation spectrum that are not
described by the Nambu--Goto action.

In the present construction, this interaction follows
from microscopic matching to the four dimensional
Abelian-Higgs theory. The pseudoscalar is dual to
the single physical polarization of the localized
massive vector mode identified in
Ref.~\cite{Alameda-Calvo:2026dat}, with its mass and
couplings determined by transverse integrals involving
the vortex background and normalized mode profiles.
The vector and pseudoscalar formulations thus describe
the same localized degree of freedom, providing an
explicit microscopic realization of the geometric
interaction employed in effective theories of
QCD flux tubes.

The topological character of the curvature density
also clarifies the axion interpretation. A constant
shift of $\vartheta$ changes the geometric coupling
only by a topological contribution, whereas the mass
term explicitly breaks the continuous shift symmetry.
The resulting theory provides a microscopic framework
for studying axion interactions and energy transfer
between transverse motion and internal string
excitations. Its geometric structure connects this theory 
to effective descriptions of QCD flux tubes. The BPS 
specific coupling relations and scalar spectrum follow from the 
present model, although similar features may arise in other 
field theories supporting flux tubes.

\section{Numerical tests of the effective action}
\label{sec:numerical-tests}

At leading order, the effective action reduces to the Nambu--Goto
description of the transverse Goldstone modes. Field theory simulations
of local strings have tested this description in several settings,
including traveling disturbances and loop evolution
\cite{Nagasawa:1994md,Olum:1999sg,Blanco-Pillado:2023sap,Helfer:2018qgv,Aurrekoetxea:2020tuw,Aurrekoetxea:2023vtp}.
Even when Nambu--Goto dynamics provides the leading approximation,
the excitation of internal modes can induce subleading corrections
to the string trajectory. Recent simulations with adaptive mesh refinement confirm that
Nambu--Goto dynamics governs circular loop collapse at leading
order, while resolving subleading corrections arising from the
backreaction of the shape excitation, accurately described by its
coupling to the intrinsic worldsheet curvature
\cite{Aurrekoetxea:2026xrl}.

Localized internal modes can also affect string dynamics
through parametric instabilities that transfer energy
from internal excitations to transverse motion. This
mechanism has been identified in domain wall
strings~\cite{Blanco-Pillado:2022rad}, axionic
strings~\cite{Blanco-Pillado:2022axf}, and local
strings~\cite{Aurrekoetxea:2026xrl,Alameda-Calvo:2026dat}.
These findings suggest that such instabilities may be
a generic feature of extended solitons supporting
bound modes localized in their cores.

For local strings, both the shape and vector modes
can amplify transverse Goldstone modes, but with
different polarization patterns. A shape excitation
amplifies a planar disturbance while preserving its
polarization, whereas a vector excitation couples
orthogonal transverse directions and produces helical
deformations. Our effective action explains this
difference through the couplings of the two internal
modes to intrinsic and normal curvature, respectively.

In this section we first examine these instabilities and 
show that their leading growth rates coincide with the 
predictions of our effective theory. We then consider the complementary process in
which colliding transverse disturbances excite the bound modes.
Finally, we isolate the direct interaction between the vector and
shape modes. Throughout these comparisons, the interaction
coefficients take the values obtained from the microscopic calculation outlined
earlier. We evolve the Abelian-Higgs equations of motion using
\texttt{GRDzhadzha}~\cite{Aurrekoetxea:2023fhl},
developed by the GRTL Collaboration (\url{www.grtlcollaboration.org}). Adaptive mesh
refinement allows us to resolve the relevant scales,
from the string core to the full string configuration.

\subsection{Parametric excitation of the Goldstone modes}
\label{sec:parametric-instabilities}

To demonstrate this instability, we consider a straight
string with a small, spatially homogeneous excitation
of one internal mode and a weak transverse perturbation
that seeds the instability. During the initial growth
stage, we neglect depletion of the homogeneous excitation
and radiation, keeping terms quadratic in the transverse
displacement $X^i(t,z)$ and linear in the internal mode
amplitude. The relevant interactions are
\begin{equation}
\label{eq:validation-geometric-interactions}
S_{\rm int}
=\frac{1}{e^2}\int d^2\sigma\,\sqrt{-\gamma}\,
\left[
\kappa_S\chi\mathcal R
-\frac{\kappa_V}{2}G_{AB}\Omega^{AB}
\right].
\end{equation}
For both internal excitations, we consider the transverse
Fourier component
\begin{equation}
\label{eq:validation-transverse-ansatz}
X^1(t,z)=X(t)\cos(kz),
\quad
X^2(t,z)=Y(t)\sin(kz),
\end{equation}
where $k=2\pi n/L_z$, with $n>0$, on a periodic interval
of length $L_z$. This ansatz allows us to compare the
evolution of the two transverse amplitudes and their
coupling to each internal mode.

For a homogeneous shape oscillation,
$\chi(t)=\chi_0\cos(\omega_b t)$, the interaction is
determined by the quadratic intrinsic curvature,
\begin{equation}
\mathcal R^{(2)}
=2\left[
(\partial_t\partial_zX^i)^2
-(\partial_t^2X^i)(\partial_z^2X^i)
\right].
\end{equation}
Substituting Eq.~\eqref{eq:validation-transverse-ansatz}
and integrating over the periodic interval gives
\begin{equation}
\int_0^{L_z}dz\,\mathcal R^{(2)}
=
\frac{L_zk^2}{2}\frac{d^2}{dt^2}
\left(X^2+Y^2\right).
\end{equation}
After two integrations by parts in time, and including
the quadratic Nambu--Goto contribution, the transverse
Lagrangian becomes
\begin{equation}
\begin{aligned}
L_{\perp,S}^{(2)}
=\frac{L_z}{4e^2}\biggl\{
&\mu\left[
\dot X^2+\dot Y^2-k^2(X^2+Y^2)
\right]
\\
&+2\kappa_Sk^2\ddot\chi\left(X^2+Y^2\right)
\biggr\},
\end{aligned}
\end{equation}
where boundary terms have been omitted. The two
amplitudes are independent and obey identical equations.
Writing $\xi=X$ or $Y$, we obtain
\begin{equation}
\label{eq:validation-shape-resonance}
\ddot\xi
+k^2\left[
1+\frac{2\kappa_S\omega_b^2\chi_0}{\mu}
\cos(\omega_b t)
\right]\xi=0.
\end{equation}
This Mathieu equation describes a periodic modulation
of the restoring force, with a principal resonance
near $2k=\omega_b$. Since the shape excitation does
not mix the transverse amplitudes, an initially planar
seed with $Y=\dot Y=0$ remains planar as it grows.
This polarization dependence follows from the
contraction of transverse indices with $\delta_{ij}$
in $\mathcal R^{(2)}$.

For a homogeneous vector oscillation,
$C_z=C(t)=C_0\cos(\omega_b t)$ and $C_t=0$, one has
$G_{tz}=\dot C$, and the interaction becomes
$\kappa_V\dot C\,\Omega_{tz}/e^2$.
The quadratic normal curvature contains the
antisymmetric contraction
\begin{equation}
\Omega_{AB}^{(2)}
=\epsilon_{ij}
(\partial_A\partial_CX^i)
(\partial_B\partial^CX^j),
\end{equation}
which couples the two transverse directions.
Using the same Fourier ansatz gives
\begin{equation}
\int_0^{L_z}dz\,\Omega_{tz}^{(2)}
=-\frac{L_zk}{2}\frac{d}{dt}
\left(\dot X\dot Y-k^2XY\right).
\end{equation}
After an integration by parts in time, the transverse
Lagrangian is
\begin{equation}
\begin{aligned}
L_{\perp,V}^{(2)}
=\frac{L_z}{4e^2}\biggl\{
&\mu\left[
\dot X^2+\dot Y^2-k^2(X^2+Y^2)
\right]
\\
&+2\kappa_Vk\ddot C
\left(\dot X\dot Y-k^2XY\right)
\biggr\}.
\end{aligned}
\end{equation}
The combinations $h_\pm=(X\pm Y)/\sqrt{2}$
diagonalize this Lagrangian, with time dependent
kinetic coefficients
\begin{equation}
a_\pm(t)=\mu\mp\kappa_Vk\omega_b^2C(t).
\end{equation}
Defining $Q_\pm=\sqrt{a_\pm/\mu}\,h_\pm$ gives
\begin{equation}
\ddot Q_\pm+
\left[
k^2-\frac{\ddot a_\pm}{2a_\pm}
+\frac{\dot a_\pm^2}{4a_\pm^2}
\right]Q_\pm=0.
\end{equation}
To first order in $C_0$, this becomes
\begin{equation}
\label{eq:validation-vector-resonance}
\ddot Q_\pm+
\left[
k^2\mp\frac{\kappa_Vk\omega_b^4C_0}{2\mu}
\cos(\omega_b t)
\right]Q_\pm=0.
\end{equation}
Both channels have a principal resonance near
$2k=\omega_b$. Their opposite modulation signs
correspond to different phases relative to the
homogeneous oscillation and give equal leading
growth rates. The corresponding spatial polarizations
are proportional to $(\cos kz,\pm\sin kz)$ and describe
helical deformations. A seed in one transverse direction
is a superposition of these channels and generically
develops an orthogonal component with a relative
spatial phase. In the dual formulation, this behavior
follows from the pseudoscalar coupling to $\mathcal T$.

To compare the growth rates, one can write both Equations \eqref{eq:validation-shape-resonance} and \eqref{eq:validation-vector-resonance} as
\begin{equation}
\label{eq:validation-mathieu}
\ddot q+\left[k^2+b\cos(\omega_b t)\right]q=0.
\end{equation}
For weak modulation, the leading Floquet exponent
at the center of the principal band, expressed in
worldsheet time $t$, is $\Gamma=|b|/(2\omega_b)$.
Equations~\eqref{eq:validation-shape-resonance}
and~\eqref{eq:validation-vector-resonance} therefore give
\begin{equation}
\label{eq:validation-resonance-growth}
\Gamma_S=\frac{|\kappa_S\chi_0|\omega_b^3}{4\mu},
\qquad
\Gamma_V=\frac{|\kappa_VC_0|\omega_b^4}{8\mu}.
\end{equation}

Thus, the parametric instabilities driven by the internal mode excitations
 generate oscillatory transverse displacements with envelopes that grow exponentially as $\sim e^{\Gamma_{S,V}t}$,

To test the EFT predictions against field theory simulations,
we initialize a straight string with a small, spatially
homogeneous excitation of either the shape or the vector mode.
The instability is seeded by a transverse displacement initially
at rest,
\begin{equation}
X(0)=X_0,\qquad
\dot X(0)=Y(0)=\dot Y(0)=0,
\end{equation}
with wavenumber $k=\omega_b/2$, corresponding to the center
of the principal resonance band at leading order.

For the phases of the homogeneous oscillations specified above,
the leading resonant solutions obtained from
Eq.~\eqref{eq:validation-mathieu} in the weak-modulation regime,
$|b|/k^2\ll1$, are
\begin{equation}
\label{eq:validation-seed-shape}
\begin{aligned}
X(t)&\simeq X_0\bigl[
\cosh(\Gamma_S t)\cos(kt)
-\sinh(\Gamma_S t)\sin(kt)
\bigr],\\
Y(t)&=0,
\end{aligned}
\end{equation}
for a homogeneous shape excitation, and
\begin{equation}
\label{eq:validation-seed-vector}
\begin{aligned}
X(t)&\simeq X_0\cosh(\Gamma_V t)\cos(kt),\\
Y(t)&\simeq X_0\sinh(\Gamma_V t)\sin(kt),
\end{aligned}
\end{equation}
for a homogeneous vector excitation. These expressions retain
the cumulative resonant growth while neglecting subleading
periodic corrections and higher order corrections to the
growth rates.

To quantify the amplification, we define the combined
transverse displacement envelope
\begin{equation}
\label{eq:validation-envelope}
\begin{aligned}
\mathrm{Env}(X,Y)
&\equiv
\left[
X^2+Y^2+\frac{\dot X^2+\dot Y^2}{k^2}
\right]^{1/2}\\
&\simeq |X_0|\sqrt{\cosh(2\Gamma t)}.
\end{aligned}
\end{equation}
The second line follows from
Eqs.~\eqref{eq:validation-seed-shape}
and~\eqref{eq:validation-seed-vector} at leading order
in $\Gamma/k$. At critical coupling, equal normalized
internal-mode amplitudes, $|\chi_0|=|C_0|$, imply
$\Gamma=\Gamma_S=\Gamma_V$, with the growth rates given
by Eq.~\eqref{eq:validation-resonance-growth}.

\begin{figure}
\centering
\href{https://youtu.be/tlT1DMFg6DY}{
\includegraphics[width=\columnwidth]
{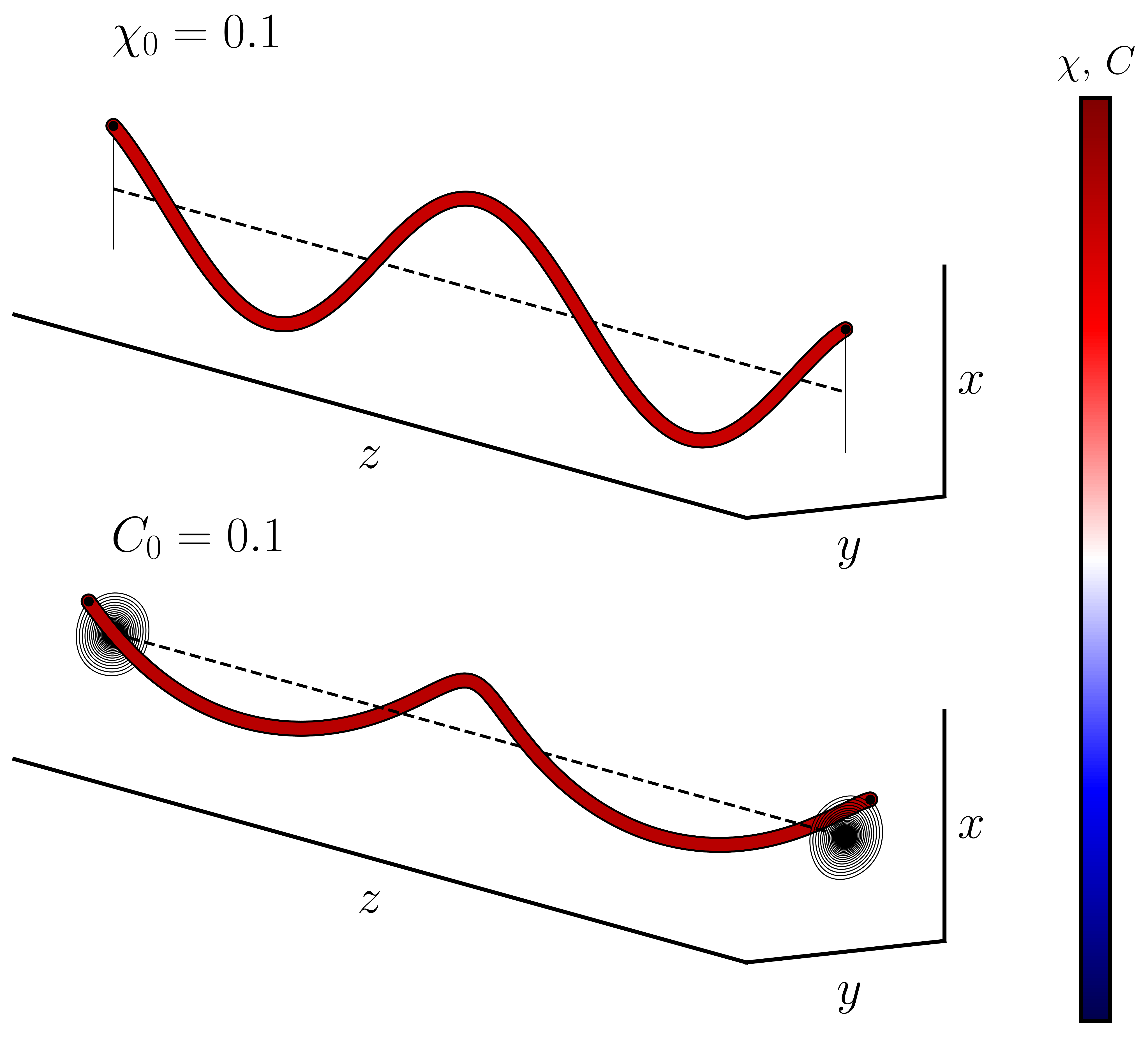}}
\caption{Transverse deformations induced by homogeneous
internal mode excitations. The upper and lower panels show
the string core for initial shape and vector amplitudes
$\chi_0=0.1$ and $C_0=0.1$, respectively.
The shape excitation preserves the seeded planar
polarization, whereas the vector excitation generates
an orthogonal transverse component, which results in a time dependent  helical configuration.
A movie is available at
\url{https://youtu.be/tlT1DMFg6DY}.}
\label{fig:validation-instability-geometry}
\end{figure}

We perform separate simulations with initial shape amplitude
$\chi_0=0.1$ and initial vector amplitude $C_0=0.1$,
using the same transverse seed amplitude $X_0=10^{-3}$.
The lattice simulations reproduce the predicted polarization
patterns, as illustrated by the string configurations in
Fig.~\eqref{fig:validation-instability-geometry}.
The shape excitation amplifies the initial planar
perturbation, while the vector excitation also generates
a displacement in the orthogonal transverse direction.

\begin{figure}
\centering
\includegraphics[width=\columnwidth]
{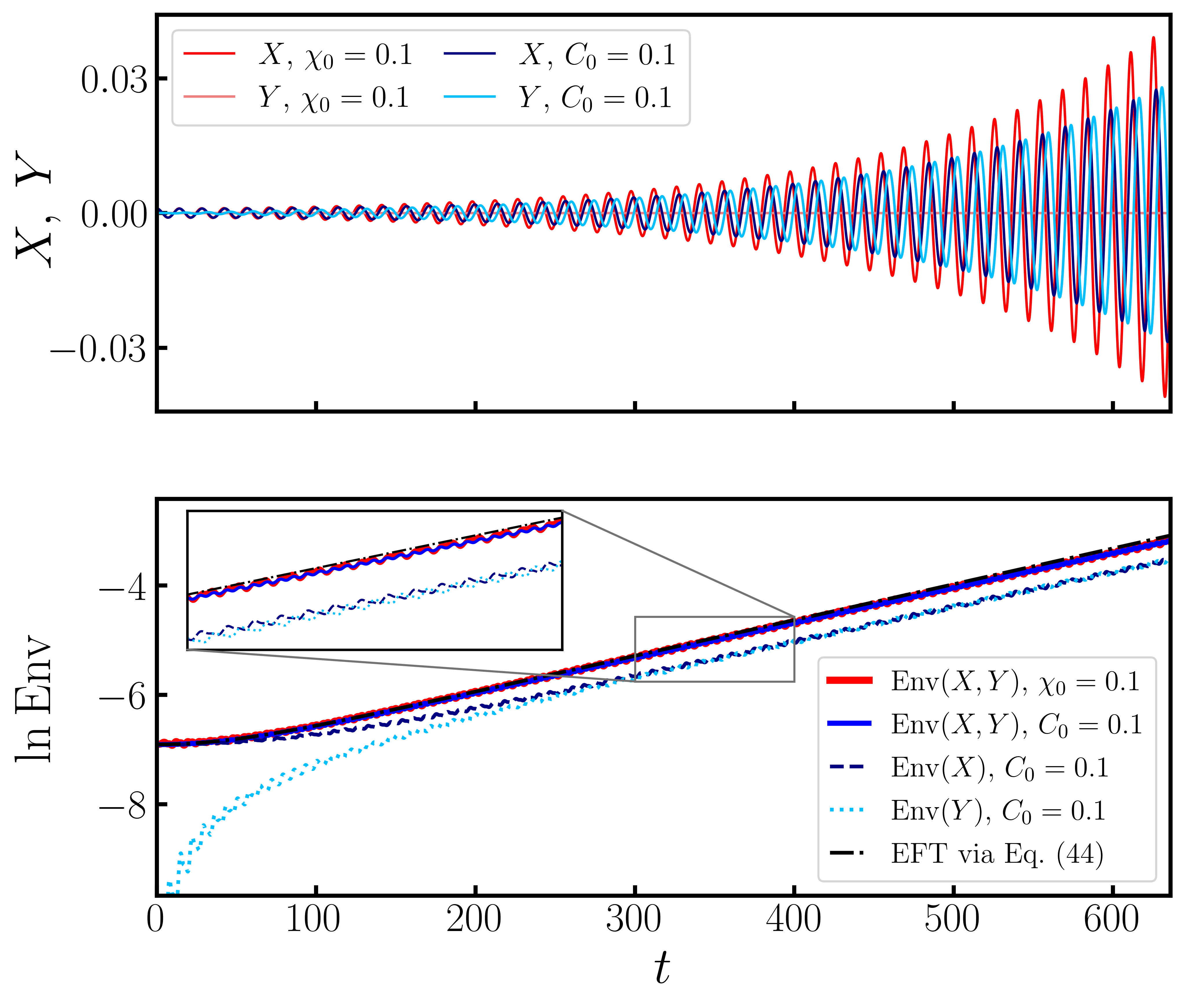}
\caption{Growth of transverse Goldstone excitations
seeded by $X_0=10^{-3}$ at $k=\omega_b/2$,
for initial shape amplitude $\chi_0=0.1$ and initial
vector amplitude $C_0=0.1$ in separate simulations.
The upper panel shows the transverse amplitudes,
while the lower panel compares the logarithms of their
combined envelopes with the leading EFT prediction
in Eq.~\eqref{eq:validation-envelope}.
For these equal normalized internal-mode amplitudes,
the BPS relation implies $\Gamma_S=\Gamma_V$,
with the common growth rate fixed by
Eq.~\eqref{eq:validation-resonance-growth}.}
\label{fig:validation-instability-growth}
\end{figure}

Fig.~\eqref{fig:validation-instability-growth} compares
the transverse amplitudes and their combined envelopes
with the leading EFT prediction in
Eq.~\eqref{eq:validation-envelope}.
The comparison uses the microscopically determined
couplings, without adjusting the predicted growth rates.
The observed agreement tests both the polarization
dependence and the strength of the geometric interactions
at the chosen resonant wavenumber. Note, however, that all comparisons are restricted to the weak excitation
regime in which the leading resonant approximation
remains applicable. 
Beyond this regime, additional interactions and the backreaction
between the transverse and internal modes must be included,
allowing for energy transfer back into the internal excitations (See Refs.~\cite{Aurrekoetxea:2026xrl,Alameda-Calvo:2026dat}
for further details.).

In Appendix~\ref{app-CC-vs-EFT}, we also show that these cubic vertices reproduce the
interactions obtained in the earlier collective
coordinate reductions~\cite{Alameda-Calvo:2026dat}. The analytic matching and
numerical comparisons thus provide independent
checks of the couplings identified in the 
covariant effective action.

\subsection{Bound mode excitation by colliding transverse waves}
\label{sec:collision-validation}

We next consider collisions of counterpropagating transverse
disturbances on a straight string, providing a complementary
test of the same geometric interactions. The incoming string
has no internal excitation, and the overlap of the two waves
generates the curvature sources that excite the bound modes.
The coefficients have already been determined by microscopic
matching and tested by the instabilities described above.
The collision experiment therefore uses the same coefficients
and tensorial couplings, with no further adjustment. Agreement 
with both numerical experiments using a single set of 
microscopically determined coefficients provides a nontrivial 
validation of the worldsheet effective theory in the regime considered.

\subsubsection{Field theory initial data and mode extraction}

We initialize each incoming disturbance using an exact
Vachaspati-Vachaspati traveling wave solution
\cite{Vachaspati:1990sk}. Each isolated null traveling wave is an exact
solution of the field theory equations. Two nonoverlapping
disturbances of compact support consequently provide exact
separated incoming data. Here we use Gaussian profiles with
an initial separation sufficiently large that their overlap
is exponentially small.

In the incoming region, the transverse displacement is
\begin{equation}
\label{eq:validation-incoming-waves}
X^i(t,z)\simeq e_+^iF(t+z)+e_-^iG(t-z),
\end{equation}
where the unit polarization vectors satisfy
$e_+\cdot e_-=\cos\delta$ and
$\epsilon_{ij}e_+^ie_-^j=\sin\delta$.
The $F$ wave moves toward decreasing $z$, and the $G$ wave
toward increasing $z$. For simplicity we take
\begin{equation}
\label{eq:validation-gaussian-profiles}
F(s)=G(s)=A\exp\left[-\frac{s^2}{2L^2}\right],
\qquad \epsilon\equiv A/L\ll1.
\end{equation}
Here $A$ is the transverse amplitude of each individual
incoming wave and $L$ is its width.
The leading collision is centered at $t=z=0$. We choose $A=0.35$ and $L=4$, so that the incoming waves are well within the perturbative regime.

We locate the string core from the zeros of the Higgs field
on each $z$ slice and extract the internal mode amplitudes by
projecting the scalar and gauge perturbations onto the
normalized straight string eigenmodes.
The shape projection includes both the Higgs and transverse
gauge contributions, while the vector projection uses the
longitudinal gauge field. The reference vortex and the
perturbations are expressed in the same gauge.
In the regular gauge used for the background, the projections
on each transverse slice are
\begin{equation}
\begin{aligned}
\chi(t,z)
&=\int d^2x\,\left[
\varphi(\rho)\,\delta|\phi|
+a(\rho)\,\delta A_{\theta}
\right],
\\
C\equiv C_z(t,z)&=\int d^2x\,\psi(\rho)A_z,
\end{aligned}
\end{equation}
where $\delta|\phi|=|\phi|-f(\rho)$ and
$\delta A_\theta\equiv A_\theta-\alpha(\rho)/\rho$ is the perturbation
of the orthonormal azimuthal component.
The mode normalizations ensure that these overlaps recover
the corresponding amplitudes at linear order. More details about this computing these quantities numerically can be found in Appendix \ref{app-convergence-test}.

\subsubsection{Geometric sources and the bound mode response}

\begin{figure}
\centering
\href{https://youtu.be/O7oJLk80Ce4}{
\includegraphics[width=\columnwidth]{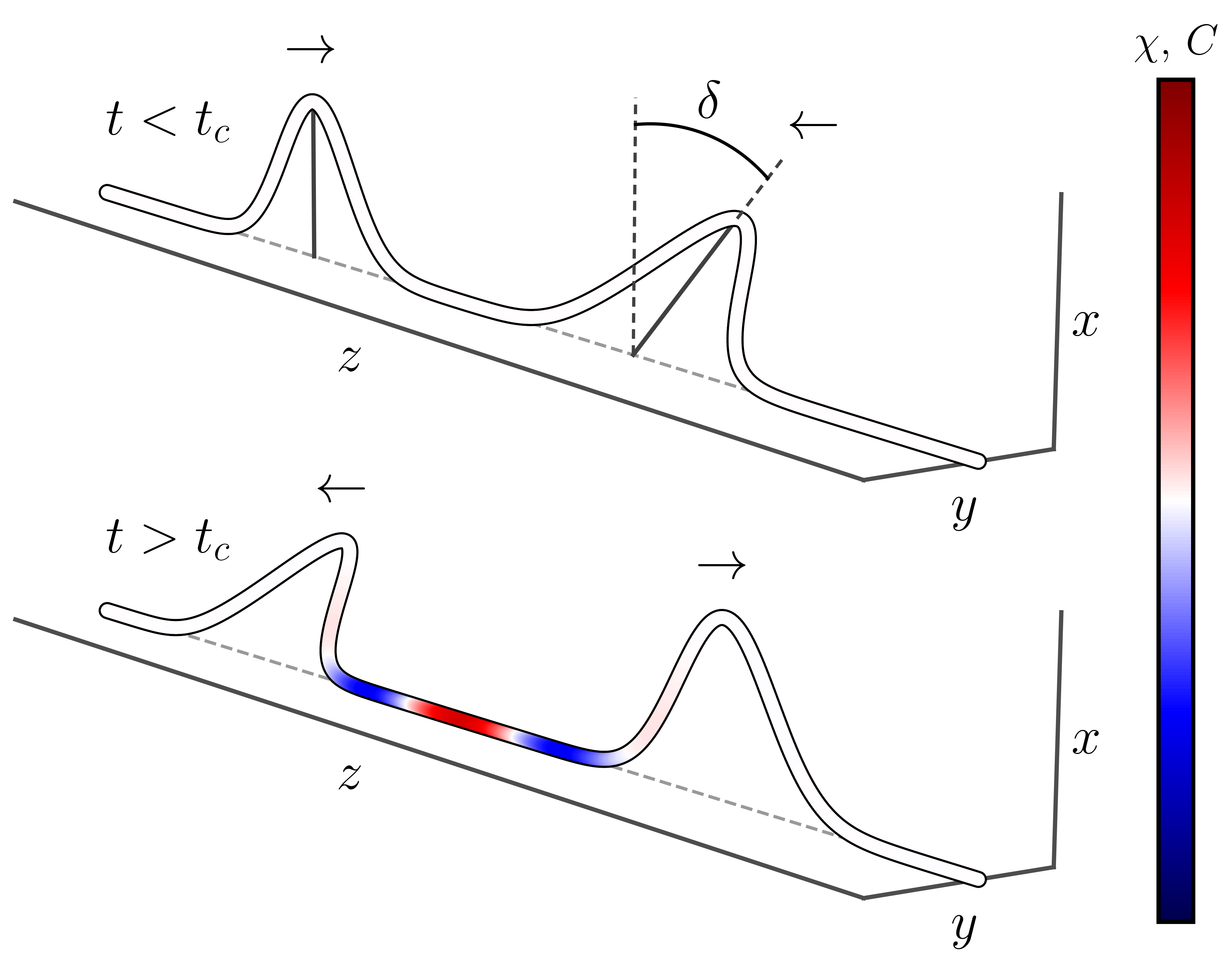}}
\caption{Excitation of the shape and vector modes by colliding
transverse waves. The upper panels show the incoming waves,
and the lower panels show the internal excitation between
the outgoing waves. The color represents the shape amplitude
for $\delta=45^\circ$, $A=0.35$, and $L=4$.
The transverse directions are stretched for visibility.
A movie is available at \url{https://youtu.be/O7oJLk80Ce4}.}
\label{fig:validation-collision-geometry}
\end{figure}

During the collision, Eq.~\eqref{eq:validation-incoming-waves}
is the leading term in the small slope expansion of the
Nambu--Goto solution. The detailed perturbative solution
and an exact solution in null coordinates are given in
Appendix~\ref{app-colliding-wiggles}.
At leading order, the induced curvature sources according to
our effective theory are,
\begin{equation}
\label{eq:validation-polarized-curvatures}
\begin{aligned}
\mathcal R&=-8F''(t+z)G''(t-z)\cos\delta+\cdots,
\\
\mathcal T&=\phantom{-}4F''(t+z)G''(t-z)\sin\delta+\cdots.
\end{aligned}
\end{equation}
An isolated traveling wave has neither source, consistently
with its exact field theory realization. The internal modes
are therefore sourced where the two incoming disturbances
overlap.

Writing $\mathcal D=\partial_t^2-\partial_z^2+\omega_b^2$,
the leading response equations are
\begin{equation}
\mathcal D\chi=\kappa_S\mathcal R,
\qquad
\mathcal D\vartheta=-\omega_b\kappa_V\mathcal T.
\end{equation}
Let $\mathcal U$ be the retarded solution of
\begin{equation}
\mathcal D\mathcal U=J(t,z),
\qquad J(t,z)=F''(t+z)G''(t-z),
\end{equation}
with no incoming massive excitation. Then, one can write the solution for
the fields as
\begin{equation}
\label{eq:validation-collision-response}
\begin{aligned}
\chi&=-8\kappa_S\mathcal U\cos\delta,
\\
\vartheta&=-4\omega_b\kappa_V\mathcal U\sin\delta.
\end{aligned}
\end{equation}
At critical coupling, $\omega_b\kappa_V=2\kappa_S$,
so the two responses have the same prefactor and
complementary cosine and sine dependences on the relative
polarization. Reversing $\delta$ leaves the scalar response
unchanged and reverses the pseudoscalar response.
This provides a direct test of the different geometric
couplings. At fixed $L$, the leading internal response is
of order $\epsilon^2$, with corrections beginning at
order $\epsilon^4$ within the truncated worldsheet theory.

For quantitative comparisons, the response can be evaluated
within the late time approximation where we assume that the 
sources have already shut down. For the Gaussian profiles, the late time response at the
collision center is
\begin{equation}
\label{eq:validation-gaussian-tail}
\mathcal U(t,0)
\simeq
\frac{\sqrt{2\pi}}{32}A^2L^2e^{-\frac{\omega_b^2L^2}{4}}\omega_b^{7/2}
\,
\frac{\sin(\omega_bt+\pi/4)}{\sqrt{t}}.
\end{equation}
The derivation and the full outgoing integral
solution are given in Appendix~\ref{app-colliding-wiggles}.
This asymptotic expression requires
$t\gg\max\{L,\omega_bL^2,\omega_b^{-1}\}$,
while decay and backreaction remain small.

The exponential factor shows that residual bound mode
production in these Gaussian collisions is suppressed
when $L$ is large compared with the inverse bound-state
mass. It measures the spectral overlap of the curvature
source with the massive worldsheet dispersion relation.
The $t^{-1/2}$ envelope describes dispersive propagation
along the string. These features are consistent with the
shape excitation found in collisions of disturbances on
domain wall strings \cite{Blanco-Pillado:2022rad}.

\begin{figure}
\centering
\includegraphics[width=\columnwidth]{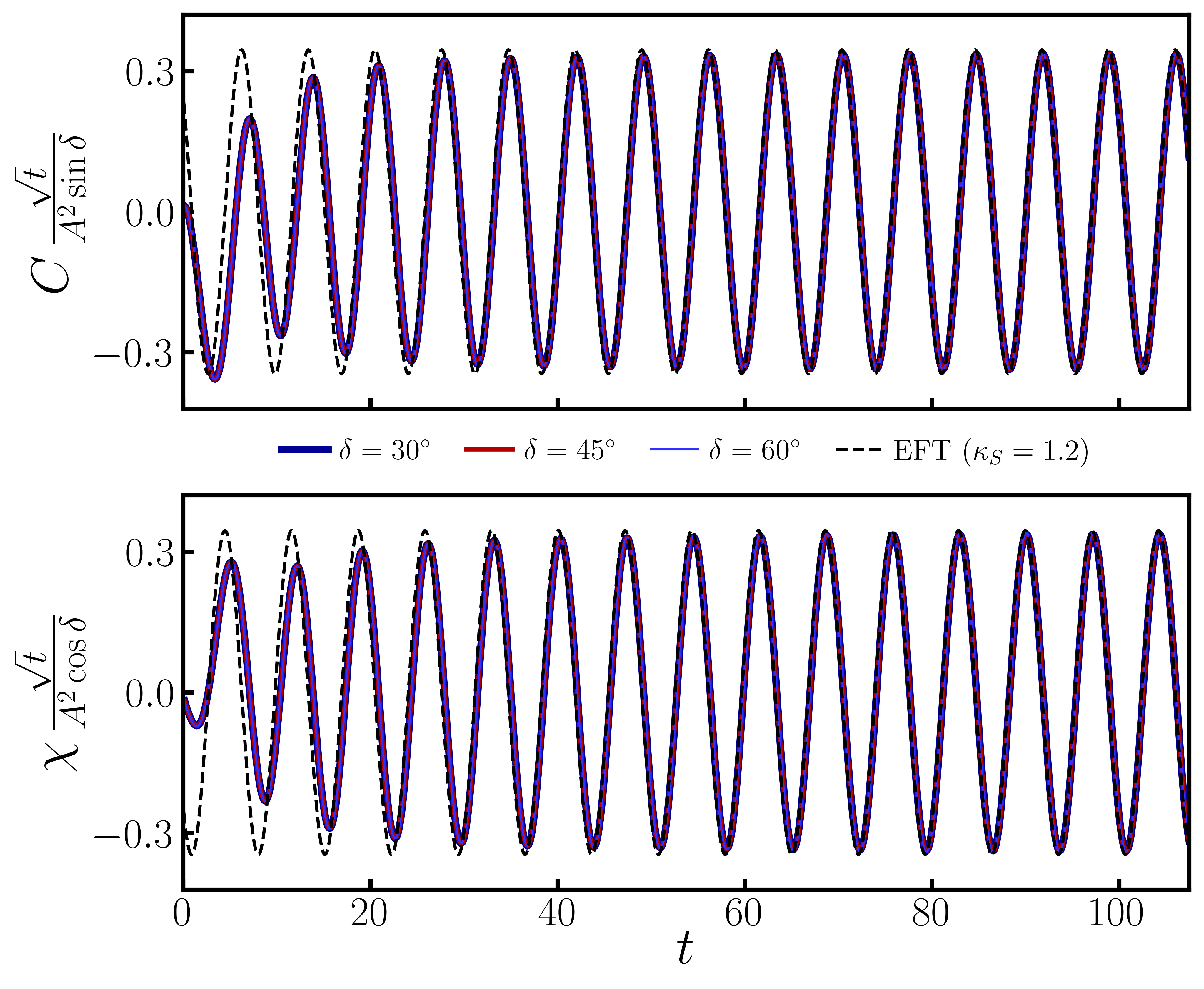}
\caption{Vector and shape amplitudes at the collision center,
compared with the effective-theory predictions. The vector
observable is $C_z$, obtained from the pseudoscalar response
through Eq.~\eqref{eq:validation-vector-observable}.
Time is measured from the collision. The displayed
$\sqrt t$ rescaling highlights the dispersive envelope.}
\label{fig:validation-collision-modes}
\end{figure}

The lattice vector observable is $C_z$. At leading order,
the duality relation gives
\begin{equation}
\label{eq:validation-vector-observable}
C_z=\frac{\partial_t\vartheta}{\omega_b},
\qquad
C_t=\frac{\partial_z\vartheta}{\omega_b}.
\end{equation}
We therefore compare the projected vector signal with
$\partial_t\vartheta/\omega_b$.
At late times, $C_z$ is in temporal quadrature with
$\vartheta$, while its leading envelope has the same
normalization.  Figure~\ref{fig:validation-collision-modes} shows excellent agreement  between the internal mode amplitudes
extracted from the field theory simulations and the effective
theory predictions. This agreement extends to both the absolute
amplitudes produced by the collision and their dependence on
the relative polarization angle $\delta$, providing a
simultaneous test of the coupling strength $\kappa_S$ and
the geometric structure of the interactions.

\subsubsection{Nambu--Goto time delay }

The preceding analysis assumes that the transverse wave
packets follow Nambu--Goto dynamics during the collision.
We now test this assumption independently by measuring
their time delay after they separate. This delay provides
a direct probe of the nonlinear interaction between the
two disturbances. The Nambu--Goto equations admit an exact
solution for the collision of counterpropagating transverse
profiles~\cite{Barbashov:1966frq,Barbashov:1966nvq}, allowing
a quantitative comparison between the predicted delay in NG
and the full field theory evolution.

The exact NG solution shows that after the waves separate, 
each retains its incoming shape and acquires a delay determined 
by the opposite profile, namely
\begin{equation}
\label{eq:ng-packet-delay}
\Delta t_G=\int_{-\infty}^{\infty}ds\,[F'(s)]^2,
\qquad
\Delta t_F=\int_{-\infty}^{\infty}ds\,[G'(s)]^2.
\end{equation}
For the Gaussian convention in
Eq.~\eqref{eq:validation-gaussian-profiles}, this gives
\begin{equation}
\label{eq:validation-gaussian-delay}
\Delta t_F=\Delta t_G=\frac{\sqrt\pi A^2}{2L}.
\end{equation}
The result is independent of $\delta$.
In physical units, the delay of each wave equals the
excess energy of the counterpropagating wave divided by
the string tension. Appendix~\ref{app-colliding-wiggles}
derives these statements both in perturbative static
gauge and from the exact null coordinate solution.

\begin{figure}
\centering
\includegraphics[width=\columnwidth]{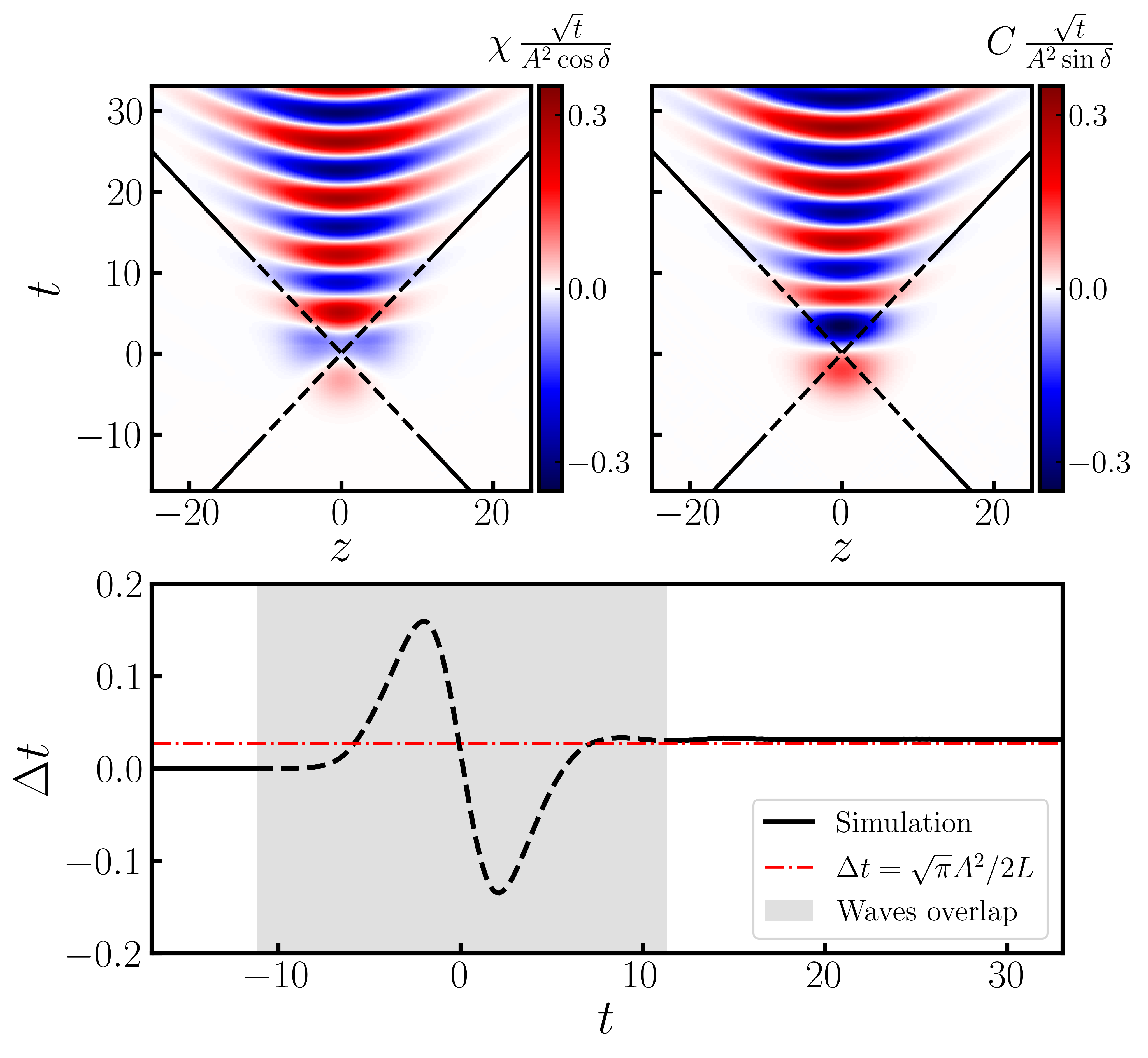}
\caption{Upper panels show the shape and vector amplitudes
in spacetime after the collision of two transverse waves.
The lower panel compares the measured outgoing delay with
Eq.~\eqref{eq:validation-gaussian-delay}.
The overlap interval is shaded, and the mode amplitudes
are rescaled by $\sqrt t$ for $t>0$.}
\label{fig:validation-spacetime-delay}
\end{figure}

We measure the delay by tracking the center $z_w(t)$
of an outgoing Gaussian disturbance, obtained by fitting
the appropriate transverse component of the core
displacement. For a wave moving toward positive $z$,
its lag relative to free propagation from an initial
time $t_i$ is
\begin{equation}
\Delta t(t)=(t-t_i)-[z_w(t)-z_w(t_i)].
\end{equation}
This lag approaches a constant after the waves separate.
The fitting procedure does not assign a reliable individual
wave position during the overlap interval.
Figure~\ref{fig:validation-spacetime-delay} shows the
agreement of the outgoing delay with the Nambu--Goto
prediction. Together with the bound mode comparison,
this supports using the leading Nambu--Goto solution
to compute the geometric sources of the internal response.

The comparison assumes small incoming slopes and times
over which nonlinear decay of the bound modes and their
backreaction remain perturbatively small. The worldsheet
description also requires negligible emission into bulk
fields, which depends on both the amplitude and the width
relative to the transverse core scale. We furthermore restrict attention
to smooth configurations for which cusp formation\footnote{In 
the Nambu--Goto description, cusps are points
where the string momentarily reaches the speed of light
and folds back on itself~\cite{Turok:1984cn,Blanco-Pillado:1998tyu,Olum:1998ag}.}and
other strongly curved regions are absent
\cite{Olum:1998ag,Olum:1999sg,Blanco-Pillado:2023sap}.
The convergence study reported in Appendix \ref{app-convergence-test}
using three resolutions, supports fourth-order convergence
of the collision induced vector signal.

For the amplitudes and widths considered, the simulations
support a description in which Nambu--Goto dynamics
governs the leading string motion and the geometric
sources excite the localized modes with subleading
backreaction. Increasing the amplitude or decreasing the
width enhances the importance of additional interactions
and finite width corrections. An appreciable bulk flux
requires the corresponding bulk degrees of freedom in
the dynamical description.

Finally, let us also briefly discuss the possibility of integrating
out the bound modes at low energies.
When characteristic frequencies are well below the bound
state masses and on-shell production is negligible,
integrating out the massive modes instead yields a local
derivative expansion. The shape mode exchange generates an
$\mathcal R^2$ interaction. On the other hand, the pseudoscalar 
exchange contributes to $\Omega_{AB}\Omega^{AB}$ and cancels 
the contact term generated by dualization at leading derivative order,
leaving the independent contribution inherited from the
vortex background.  Other terms omitted from our truncation may contribute at the same derivative order. 

However, the production and parametric amplification
processes studied here require explicit dynamical internal
fields.

\subsection{Direct coupling between the vector and shape modes}
\label{sec:direct-mode-coupling}

Finally, the interaction $g_{SV}\chi C_A C^A$ can be
isolated in a straight string simulation. We initialize
a small, uniform vector excitation and choose the
periodic length $L_z$ so that the allowed nonzero
wavenumbers $2\pi n/L_z$ lie outside the principal
vector induced resonance band
\cite{Alameda-Calvo:2026dat}.
This suppresses transverse growth during the comparison,
allowing the induced shape excitation to be identified
more clearly.

For $C_z\simeq C_0\cos(\omega_b t)$, $C_t=0$, and
$\chi(0)=\dot\chi(0)=0$, the leading prediction for
the shape mode is described by the equation,
\begin{equation}
\ddot\chi+\omega_b^2\chi
=-g_{SV}C_0^2\cos^2(\omega_b t),
\end{equation}
which can be solved exactly at this order to yield,
\begin{equation}
\label{eq:validation-vector-to-shape}
\chi(t)
=\frac{g_{SV}C_0^2}{6\omega_b^2}
\left[\cos(2\omega_b t)+2\cos(\omega_b t)-3\right].
\end{equation}

The forcing contains a constant part and a second
harmonic. Since the induced shape amplitude is
$\mathcal O(C_0^2)$, the scalar self-interaction
$3/2g_{SV}\chi^2$ first contributes at order $C_0^4$ to
its equation of motion so we neglect it here.

\begin{figure}
\centering
\includegraphics[width=\columnwidth]{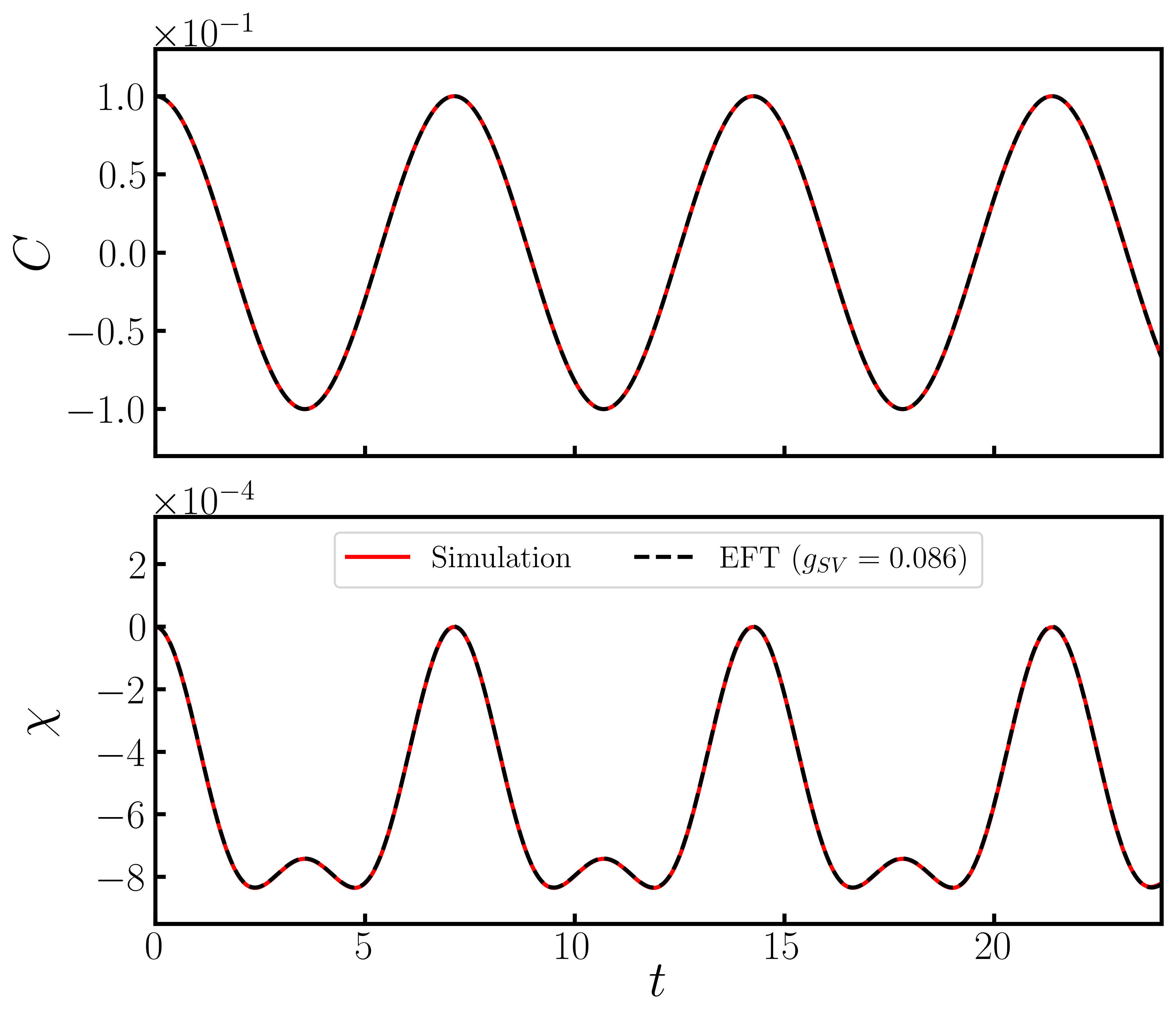}
\caption{Shape excitation induced by a homogeneous vector
mode with initial amplitude $C_0=0.1$, compared with
Eq.~\eqref{eq:validation-vector-to-shape} using the
microscopically determined value of $g_{SV}$, Eq. \eqref{eq:eft-matched-coefficients}.}
\label{fig:validation-vector-shape}
\end{figure}

Figure~\ref{fig:validation-vector-shape} compares the
projected field theory amplitudes with this prediction.
The amplitude and harmonic content test the direct mixed
interaction, independently of the geometric couplings.
The effective theory also predicts quadratic scaling of
the induced shape amplitude with $C_0$. The agreement
between the effective action and the field theory results
is remarkable.

Together with the instability and collision experiments,
this comparison tests energy exchange among the
translational, shape, and vector degrees of freedom
using one set of microscopically determined coefficients.

\section{Conclusions}

We have derived a covariant worldsheet action that connects the
microscopic structure of Abelian-Higgs strings to their dynamics
beyond the Nambu--Goto approximation. At critical coupling, the
theory includes the transverse Goldstone modes, a massive scalar
shape mode, and a massive vector mode, together with their
geometric and direct interactions. All coefficients retained at
this order are determined by transverse integrals of the vortex
background and normalized mode profiles. The resulting description
is therefore predictive, with no additional phenomenological
parameters at this order in the effective action expansion.

The dynamics has a simple geometric interpretation. The shape mode
couples to intrinsic worldsheet curvature, while the pseudoscalar
dual to the vector couples to the curvature of the normal bundle.
These interactions explain why shape excitation amplifies planar
transverse motion, whereas vector excitation produces helical
deformations. Despite their different polarization structures,
the two instabilities have equal leading growth rates
for equal normalized internal amplitudes at critical coupling.
The action also identifies the direct interaction responsible
for vector induced shape excitation and reproduces the earlier
collective coordinate description through cubic order.

Lattice field theory simulations test this description through
three complementary processes: parametric amplification of
transverse motion, excitation of the internal modes by colliding
transverse waves, and direct energy transfer between the vector
and shape modes. The same microscopically determined coefficients
describe all three processes without further adjustment.
In particular, the scattering experiments reproduce the predicted
complementary dependence of the internal response on the relative
polarization of the incoming waves. Their collision time delay
also agrees with the Nambu--Goto prediction, supporting a
description in which the leading string trajectory supplies the
geometric sources for the internal excitations. Together, these
comparisons provide nontrivial tests of both the interaction
strengths and their geometric structure in the regime studied.

The construction also suggests applications beyond cosmic strings.
The dual pseudoscalar has the same geometric interaction as the
worldsheet axion proposed for confining QCD flux tubes
\cite{Dubovsky:2013gi}, providing an explicit realization of this
coupling through a localized gauge excitation in a four dimensional
field theory. More generally, the separation of transverse motion,
internal excitations, and their geometric couplings offers a useful
starting point for studying other flux tubes and vortex lines,
including those in condensed matter. Their spectra and symmetries
will differ, but the microscopic matching strategy provides a
common way to investigate how internal structure affects their
dynamics.

For cosmology, these results provide a first step toward
incorporating the effects of internal string excitations into
future studies of network evolution beyond the Nambu--Goto
description. Energy transfer into these modes
constitutes an additional dynamical channel that may contribute
to the discrepancies between Nambu--Goto and field theory simulations
of string networks as suggested in~\cite{Hindmarsh:2021mnl}. On the other
hand, the exponential suppression of residual excitation
in wide Gaussian collisions suggests that smooth, slowly
varying string segments are inefficient at exciting the
bound modes through this mechanism. Appreciable production
requires geometric sources with spectral support on the
massive worldsheet spectrum, whose threshold lies close to
that of the bulk continuum in the present model. This suggests
that the production of worldsheet excitations may be concentrated
in localized regions undergoing rapid deformation. Such conditions
may arise in regions of large curvature, including those observed
in the field theory simulations of
Ref.~\cite{Blanco-Pillado:2023sap}. 

In a cosmological network, intercommutation may provide
an additional source of internal excitations. Reconnection
between distinct strings, or between separated segments
of the same string, involves a rapid rearrangement of
the vortex cores that may populate the localized bound
modes. Although local in spacetime, this process couples
distinct worldsheets or intrinsically separated points
on the same worldsheet and changes string connectivity.
It is therefore nonlocal from the perspective of the
worldsheet description used here and is not captured
by the present effective action. This excitation mechanism 
may be related to the energy exchange between translational motion and internal modes
studied in multivortex collisions~\cite{Krusch:2024vuy, Alonso-Izquierdo:2026icz, AlonsoIzquierdo:2026mub}.
Similar studies of string intercommutation could clarify
the role of reconnection in exciting localized internal
modes.

These configurations provide
a natural setting for investigating whether internal excitations
can influence string evolution and its cosmological signatures.
Establishing their role will require extending the present
analysis beyond the smooth configurations studied here.

In this work, we have focused on bound modes localized on the
string worldsheet. The adapted coordinate construction also
provides a systematic route to incorporating radiation by retaining
the relevant bulk fluctuations and deriving their couplings to
the string and its internal modes. This extension would strengthen
the analogy with worldline effective theories of compact objects,
providing a unified description of conservative worldsheet dynamics,
resonant excitation of localized degrees of freedom, and dissipation
through radiation into the bulk.

The same techniques could also be adapted to global strings to include
their internal bound modes~\cite{Blanco-Pillado:2021jad,Blanco-Pillado:2022axf} and their couplings to the
standard effective action of the global string~\cite{Kalb:1974yc,Vilenkin:1986ku,Davis:1988rw}.
(See also some recent results on this topic in \cite{Kaltschmidt:2026hpj,Battye:2026iuy}.)

These extensions would help establish when microscopic
string structure has observable consequences for cosmological
evolution. Work in this direction is underway.

\acknowledgments
We are grateful to Maximilian Bachmaier and Andrzej Wereszczynski for stimulating discussions and comments on previous versions of the manuscript. 
We would also like to thank Katy Clough, Thomas Helfer, Eugene Lim, Dina Traykova and the GRTL collaboration (\url{www.grtlcollaboration.org})
for their support and code development work. This work has been supported in part by the PID2024-156016NB-I00 grants funded by MCIN/AEI/10.13039/501100011033/and by ERDF; ``A way of making Europe''; the Basque Government grant (IT-1977-26) and the Basque Foundation for Science (IKERBASQUE). This publication is part of the R\&D\&i project PID2025-172448NB-I00, funded by
MICIU/AEI/10.13039/501100011033. AGMC acknowledges support from the Xunta de Galicia (Consellería de Educación, Ciencia, Universidades e Formación Profesional) through Grant No.~ED431B-2024/42 and the postdoctoral fellowship program [Grant No.~ED481B-2025/059].  JCA acknowledges funding from the Department of Physics at MIT through a CTP Postdoctoral Fellowship. JQ has been supported in part by the Spanish Ministerio de Ciencia e Innovación (MCIN) with funding from the grant PID2023-148409NB-I00 MTM.

This work used Stampede3 at the Texas Advanced Computing Center through allocations PHY260066 and  PHY090003 from the Advanced Cyberinfrastructure Coordination Ecosystem: Services \& Support (ACCESS) program \cite{ACCESS}, which is supported by U.S. National Science Foundation grants 2138259, 2138286, 2138307, 2137603, and 2138296. The authors acknowledge the Texas Advanced Computing Center (TACC) at The University of Texas at Austin for providing computational resources that have contributed to the research results reported within this paper. URL: \url{www.tacc.utexas.edu}.

We acknowledge the use of Claude (by Anthropic) for assistance with text editing, checking of mathematical derivations, and the development of analysis and plotting scripts. The authors reviewed all such material and take full responsibility for the content of this work.

\appendix

\section{Computing the Effective Action from the top down}
\label{app-Effective-Action-Computation}

\subsection{General procedure and expansion of the action}

We obtain the worldsheet action by expressing the microscopic
fields in coordinates adapted to a prescribed string embedding,
expanding in the amplitudes of the retained bound modes, and
integrating over the transverse coordinates. In the dimensionless
variables of the main text, the Abelian-Higgs action at critical
coupling is

\begin{align}
\notag S=-\frac1{e^2}\int d^2\sigma\,d^2u\,\sqrt{-g}\,
\biggl[&\frac14F_{\alpha\beta}F^{\alpha\beta}
+\frac12(D_\alpha\phi)^*D^\alpha\phi\\
&+\frac18\bigl(|\phi|^2-1\bigr)^2\biggr],
\label{eq:eft11-microscopic}
\end{align}

where $D_\alpha=\partial_\alpha-iA_\alpha$ and the spacetime
signature is $(-,+,+,+)$. Greek indices refer to all four adapted
coordinates, $A,B=\lbrace{0,1\rbrace}$ to the worldsheet, and $i,j=\lbrace{1,2\rbrace}$ to the
normal directions.

The adapted coordinates are
\begin{equation}
\label{eq:eft11-coordinates}
y^\mu(\sigma,u)=X^\mu(\sigma)+u^i n_i^\mu(\sigma),
\qquad \rho^2=\delta_{ij}u^iu^j.
\end{equation}
where $\sigma^A$ are the coordinates on the worldsheet and $e_A^\mu=\partial_A X^\mu$,
$\gamma_{AB}=e_A^\mu e_{B\mu}$, and $n_i^\mu$ are orthonormal
spacelike normals. We use
$K_{iAB}=-n_{i\mu}\partial_Ae_B^\mu$,
$K_i=\gamma^{AB}K_{iAB}$, and
$\omega_A=n_{1\mu}\partial_A n_2^\mu$, with
\begin{equation}
\begin{aligned}
\label{eq:eft11-invariants}
\mathcal R&=K_iK_i-K_{iAB}K_i{}^{AB},\\
\Omega_{AB}&=\nabla_A\omega_B-\nabla_B\omega_A.
\end{aligned}
\end{equation}
Indices on worldsheet fields are raised with $\gamma^{AB}$ and
normal indices with $\delta^{ij}$. Components of microscopic
fields are instead raised with the full
inverse metric $g^{\alpha\beta}$. The construction applies within
a nonsingular tubular neighborhood, with curvature radii large
compared with the transverse extent of the profiles of the fields
we consider.

We place the straight string profiles in each local normal plane.
In the gauge where the Higgs phase is $\theta(u)$, this gives
\begin{equation}
\begin{aligned}
\label{eq:eft11-background}
\bar\phi&=f(\rho)e^{i\theta},\\
\bar A_i&=\alpha(\rho)\partial_i\theta,\\
\bar A_A&=[1-\alpha(\rho)]\omega_A~,
\end{aligned}
\end{equation}
where a overbar denotes evaluation on this background, with the
internal amplitudes set to zero. In particular,
$\bar F_{\alpha\beta}=\partial_\alpha\bar A_\beta-
\partial_\beta\bar A_\alpha$ and
$\bar D_\alpha=\partial_\alpha-i\bar A_\alpha$.
The component $\bar A_A$ accounts for the rotation of the normal
frame and is required by the accompanying gauge covariance.
It is present even when the independent vector excitation vanishes.

We then set the total fields as the sum of these background fields
plus the internal modes, namely, $\phi=\bar\phi+\delta\phi$ and
$A_\alpha=\bar A_\alpha+\delta A_\alpha$, with
\begin{equation}
\begin{aligned}
\delta\phi
&=\chi(\sigma)\varphi(\rho)e^{i\theta},
\\
\delta A_i
&=\rho a(\rho)\chi(\sigma)\partial_i\theta,
\\
\delta A_A
&=-\rho a(\rho)\chi(\sigma)\omega_A
+\psi(\rho)C_A(\sigma).
\end{aligned}
\end{equation}
Here the real functions $\varphi(\rho)$ and
$a(\rho)$ specify the radial profiles of the Higgs and
azimuthal gauge perturbations that together constitute
the shape mode. These profiles are determined by the
linearized fluctuation equations about the straight
vortex. Their common coefficient $\chi(\sigma)$ is
a real worldsheet scalar that describes the local
amplitude and propagation of this internal excitation.

Similarly, $\psi(\rho)$ specifies the transverse
localization of the vector mode, while $C_A(\sigma)$
describes its amplitude and dynamics along the
worldsheet. Thus, the radial profiles encode the
microscopic structure of the bound modes, whereas
$\chi$ and $C_A$ are the dynamical fields of the
worldsheet effective theory.

The term $-\rho a(\rho)\chi(\sigma)\omega_A$ is the
geometric completion of the shape perturbation, where 
$\omega_A$ is the normal bundle connection introduced earlier,
which describes the rotation of the local normal
basis along the worldsheet. This contribution ensures
that the shape mode ansatz transforms consistently
under local rotations of that basis.
The worldsheet field strength associated with the
vector excitation is
\begin{equation}
G_{AB}
=\nabla_A C_B-\nabla_B C_A.
\end{equation}

Using this approach, our action can be obtained using the following 
formal Taylor expansion
\begin{equation}
\label{eq:eft11-taylor}
S=\sum_{n=0}^{\infty}S^{(n)},\qquad
S^{(n)}=\left.\frac1{n!}\delta^nS\right|_{\bar\varphi,\bar A}.
\end{equation}
The superscript counts powers of $\chi$ and $C_A$. The background solves the
straight string equations, but generally not the full field
equations for a curved embedding. Its first variation can therefore
be nonzero and provides geometric sources for the internal modes.
The contributions given in the main text action are
\begin{equation}
\label{eq:eft11-organized-expansion}
\begin{aligned}
S_{\rm eff}={}&S^{(0)}+S_S^{(1)}+S_V^{(1)}\\
&+S_S^{(2)}+S_V^{(2)}+S_{SV}^{(3)}+\cdots.
\end{aligned}
\end{equation}
Here $S$ and $V$ label the shape and vector modes. We calculate their
leading kinetic and mass terms, their leading linear geometric
couplings, and the direct mixed cubic interaction. The ellipsis
includes further interactions and curvature corrections, including
mixed terms quadratic in the internal amplitudes. All these
contributions are computed separately below.

For radial transverse integrands, the geometric reduction follows
from
\begin{equation}
\label{eq:eft11-measure}
\begin{aligned}
\sqrt{-g}=\sqrt{-\gamma}\biggl[1+u^iK_i
+\frac12u^iu^j
\bigl(K_iK_j-K_{iAB}K_j{}^{AB}\bigr)\biggr].
\end{aligned}
\end{equation}
Using $\langle u^iu^j\rangle_\theta=\rho^2\delta^{ij}/2$ gives
\begin{equation}
\label{eq:eft11-angular-measure}
\frac1{2\pi}\int_0^{2\pi}d\theta\,\sqrt{-g}
=\sqrt{-\gamma}\left(1+\frac{\rho^2}{4}\mathcal R\right).
\end{equation}
This factor fixes both Ricci couplings. In the leading quadratic
operators we also use $g^{AB}=\gamma^{AB}+O(uK)$; odd transverse
moments vanish, and the remaining curvature corrections belong
to the ellipsis. Finally, we will consider that localized perturbations 
are regular at the origin and decay at infinity. 

\subsection{Zeroth order: tension and Euler term}

The transverse Higgs gradient, magnetic Maxwell term, and Higgs
potential in Eq.~\eqref{eq:eft11-microscopic} give, respectively,
\begin{equation}
\label{eq:eft11-zero-microscopic}
\begin{aligned}
S^{(0)}={}&-\frac1{e^2}\int d^2\sigma\,d^2u\,\sqrt{-g}\\
&\times\biggl[\frac12 f'^2+\frac{(1-\alpha)^2f^2}{2\rho^2}
+\frac{\alpha'^2}{2\rho^2}\\
&\hspace{3.1em}
+\frac18(f^2-1)^2\biggr]+\cdots.
\end{aligned}
\end{equation}
where primes denote $d/d\rho$. The BPS equations and boundary conditions
are
\begin{equation}
\label{eq:eft11-bps-profiles}
f'=\frac{1-\alpha}{\rho}f,\qquad
\alpha'=\frac{\rho}{2}(1-f^2),
\end{equation}
with $f(0)=\alpha(0)=0$ and $f(\infty)=\alpha(\infty)=1$.
They reduce the bracket in Eq.~\eqref{eq:eft11-zero-microscopic}
to $f'^2+\alpha'^2/\rho^2$. Integrating with
Eq.~\eqref{eq:eft11-angular-measure} yields
\begin{equation}
\label{eq:eft11-zero}
S^{(0)}=-\frac1{e^2}\int d^2\sigma\sqrt{-\gamma}\,
\left(\mu+\nu\mathcal R\right)+\cdots,
\end{equation}
where
\begin{equation}
\label{eq:eft11-mu-nu}
\begin{aligned}
\mu&=2\pi\int_0^\infty d\rho\,
\left(\rho f'^2+\frac{\alpha'^2}{\rho}\right)=\pi,\\
\nu&=\frac{\pi}{2}\int_0^\infty d\rho\,
\left(\rho^3f'^2+\rho\alpha'^2\right)=\frac{\pi}{2}.
\end{aligned}
\end{equation}
Both evaluations follow from
\begin{equation}
\label{eq:eft11-bps-energy}
\rho\left(f'^2+\frac{\alpha'^2}{\rho^2}\right)
=\frac12\frac{d}{d\rho}
\left[\alpha+(1-\alpha)f^2\right].
\end{equation}
For the second moment, write the bracket as
$1-(1-\alpha)(1-f^2)$ before integrating by parts. This gives
$\nu=\pi\int_0^\infty(1-\alpha)\alpha'\,d\rho=\pi/2$.
Thus $\mu$ gives the Nambu--Goto tension, equal to $\pi\eta^2$
in physical units, whereas the constant coefficient Euler term
does not affect the worldsheet equations of motion.

There is also a tangential Maxwell contribution from
$\bar F_{AB}=(1-\alpha)\Omega_{AB}$. Its leading reduction is
\begin{equation}
\label{eq:eft11-background-normal-curvature}
S^{(0)} \supset \frac{\kappa_{\Omega}}{e^2}
\int d^2\sigma\sqrt{-\gamma}\,\Omega_{AB}\Omega^{AB}.
\end{equation}
where
\begin{equation}
\kappa_{\Omega}=-\frac{\pi}{2}
\left[\int_0^\infty\rho\,d\rho\,(1-\alpha)^2\right].
\end{equation}
One can easily show that this term is fourth order in extrinsic 
curvature and belongs to the ellipsis in Eq.~\eqref{eq:eft11-zero}. We keep it 
here and in the action in the main text, however,
to compare with other similar terms that appear in 
the subsequent dualization derivations. This term 
was first introduced in the literature in \cite{Anderson:1995hv}.

\subsection{First order in the shape mode}

The linear shape term receives contributions from all three
microscopic terms in Eq.~\eqref{eq:eft11-microscopic}. Their
transverse parts give
\begin{align}
\label{eq:eft11-shape-linear-integrand}
S_S^{(1)}={}&-\frac1{e^2}\int d^2\sigma\,d^2u\,\sqrt{-g}\,\chi\notag\\
&\times\biggl[f'\varphi'+\frac{(1-\alpha)^2}{\rho^2}f\varphi
-\frac{1-\alpha}{\rho}f^2a\\
&\hspace{3.1em}+\frac{\alpha'}{\rho}\left(a'+\frac a\rho\right)
+\frac12(f^2-1)f\varphi\biggr]+\cdots.\notag
\end{align}
The first three terms arise from the Higgs gradient, the fourth
from the magnetic energy, and the last from the Higgs potential.
The straight string equations combine this bracket into a total
radial derivative:
\begin{equation}
\label{eq:eft11-shape-first-variation}
S_S^{(1)}=-\frac1{e^2}\int d^2\sigma\,d^2u\,
\frac{\sqrt{-g}}{\rho}\,\chi\,
\frac{d}{d\rho}\left(\rho f'\varphi+\alpha'a\right)+\cdots.
\end{equation}
With a flat transverse measure this is a vanishing boundary term.
For a curved worldsheet, Eq.~\eqref{eq:eft11-angular-measure}
introduces the weight $\rho^2\mathcal R/4$. One radial integration
by parts then gives
\begin{equation}
\label{eq:eft11-shape-tadpole}
\begin{aligned}
S_S^{(1)}&=\frac{\kappa_S}{e^2}
\int d^2\sigma\sqrt{-\gamma}\,\chi\mathcal R+\cdots,\\
\kappa_S&=\pi\int_0^\infty d\rho\,
\left(\rho^2 f'\varphi+\rho\alpha'a\right).
\end{aligned}
\end{equation}
The leading scalar source thus comes from
the curvature dependence of the transverse volume element.

\subsection{First order in the vector mode}

For this variation, $\delta\phi=\delta A_i=0$ and
$\delta A_A=\psi C_A$. The relevant microscopic contribution is
the Maxwell cross term
\begin{equation}
\label{eq:eft11-vector-linear-maxwell}
S_V^{(1)}=-\frac1{2e^2}\int d^2\sigma\,d^2u\,
\sqrt{-g}\,\bar F^{\alpha\beta}\delta F_{\alpha\beta}.
\end{equation}
To see why the Higgs kinetic term does not contribute, it is
essential to keep the mixed inverse metric. With
$\epsilon_{12}=1$ and $\xi^i=\epsilon_j{}^i u^j$, one has
\begin{equation}
\label{eq:eft11-mixed-metric}
g_{Ai}=-\omega_A\xi_i,\qquad
g^{Ai}=g^{AB}\omega_B\xi^i,\qquad
\xi^i\partial_i\theta=1.
\end{equation}
Together with Eq.~\eqref{eq:eft11-background}, these relations give
\begin{equation}
\label{eq:eft11-raised-background}
\begin{aligned}
\bar D^B\bar\phi&=0,\\
\bar F^{AB}&=(1-\alpha)g^{AC}g^{BD}\Omega_{CD},\\
\bar F^{iB}&=\xi^i\omega_A\bar F^{AB}.
\end{aligned}
\end{equation}
The first identity eliminates the linear Higgs current. Moreover,
$\delta F_{AB}=\psi G_{AB}$ and
$\delta F_{iA}=\psi' u_i C_A/\rho$. Since $\xi^iu_i=0$, the mixed
Maxwell contraction vanishes. The remaining term is therefore
\begin{equation}
\label{eq:eft11-vector-tadpole}
\begin{aligned}
S_V^{(1)}
&=-\frac1{2e^2}\int d^2\sigma\,d^2u\,
\sqrt{-g}\,\psi G_{AB}\bar F^{AB}\\
&=-\frac{\kappa_V}{2e^2}\int d^2\sigma\sqrt{-\gamma}\,
G_{AB}\Omega^{AB}+\cdots,\\
\kappa_V&=2\pi\int_0^\infty\rho\,d\rho\,(1-\alpha)\psi.
\end{aligned}
\end{equation}
The factor $1/2$ is the cross-term coefficient in the expansion of
$F_{\alpha\beta}F^{\alpha\beta}/4$. The source depends on the
normal curvature.

\subsection{Quadratic action for the shape mode}

Worldsheet derivatives of $\chi$ enter through the Higgs kinetic term and
the mixed Maxwell components. Combining the coordinate components
with the inverse metric gives
\begin{equation}
\label{eq:eft11-shape-kinetic}
S_S^{(2)}\big|_{\nabla\chi}
=-\frac1{2e^2}\int d^2\sigma\,d^2u\,\sqrt{-g}\,
g^{AB}(\varphi^2+a^2)\partial_A\chi\partial_B\chi.
\end{equation}
The contributions $\varphi^2$ and $a^2$ come from the Higgs and
Maxwell terms, respectively. At leading order in the curvature
expansion, we replace
$\sqrt{-g}\,g^{AB}$ by $\sqrt{-\gamma}\,\gamma^{AB}$.
The normalization
\begin{equation}
\label{eq:eft11-normalizations}
2\pi\int_0^\infty\rho\,d\rho\,\psi^2
=2\pi\int_0^\infty\rho\,d\rho\,(\varphi^2+a^2)=1
\end{equation}
then gives the canonical scalar kinetic term.

The transverse gradients and potential determine the mass.
A convenient way to obtain all their relative factors is
to complete the transverse energy into BPS squares before
expanding. At leading order in curvature, the coefficient
of $\chi^2/2$ becomes
\begin{equation}
\label{eq:eft11-shape-mass-integral}
\begin{aligned}
2\pi\int_0^\infty\rho\,d\rho\,\biggl[&
\left(\varphi'-\frac{1-\alpha}{\rho}\varphi+fa\right)^2\\
&+\left(a'+\frac a\rho+f\varphi\right)^2\biggr].
\end{aligned}
\end{equation}
The radial boundary term from the BPS completion does not
contribute, since the perturbations preserve the vortex
boundary conditions.

The BPS bound modes satisfy
\begin{equation}
\label{eq:eft11-profile-equations}
\begin{gathered}
-\psi''-\frac{\psi'}{\rho}+f^2\psi=\omega_b^2\psi,\\
\varphi=\frac{f\psi}{\omega_b},\qquad
a=-\frac{\psi'}{\omega_b}.
\end{gathered}
\end{equation}
Together with the background BPS equation
$f'=(1-\alpha)f/\rho$, these relations imply
\begin{equation}
\label{eq:eft11-shape-factorization}
\begin{aligned}
\varphi'-\frac{1-\alpha}{\rho}\varphi+fa&=0,\\
a'+\frac a\rho+f\varphi&=\omega_b\psi.
\end{aligned}
\end{equation}
Substituting these identities into
Eq.~\eqref{eq:eft11-shape-mass-integral}, the first square
vanishes and the second becomes $\omega_b^2\psi^2$.
The coefficient of $\chi^2/2$ is therefore
\begin{equation}
\label{eq:eft11-shape-mass-reduction}
2\pi\int_0^\infty\rho\,d\rho\,\omega_b^2\psi^2
=\omega_b^2
\left(2\pi\int_0^\infty\rho\,d\rho\,\psi^2\right)
=\omega_b^2,
\end{equation}
where the final equality follows from
Eq.~\eqref{eq:eft11-normalizations}.
Thus, with the kinetic term canonically normalized,
the worldsheet scalar has mass $\omega_b$.

The resulting quadratic terms in the action are
\begin{equation}
\label{eq:eft11-shape-quadratic}
S_S^{(2)}=-\frac1{2e^2}\int d^2\sigma\sqrt{-\gamma}\,
\left(\nabla_A\chi\nabla^A\chi+\omega_b^2\chi^2\right)
+\cdots,
\end{equation}
where the ellipsis denotes curvature dependent corrections
to the quadratic action.

\subsection{Quadratic action for the vector mode}

The worldsheet Maxwell components provide the vector kinetic term,
the mixed Maxwell components provide a radial gradient contribution
to its mass, and the Higgs kinetic term supplies the remaining mass
contribution. At leading order in curvature these are
\begin{equation}
\label{eq:eft11-vector-quadratic-unreduced}
\begin{aligned}
S_V^{(2)}={}&-\frac{2\pi}{e^2}\int d^2\sigma\sqrt{-\gamma}\int_0^\infty\rho\,d\rho\,
\biggl[\frac14\psi^2 G_{AB}G^{AB}\\
&\hspace{4.5em}+\frac12\bigl(\psi'^2+f^2\psi^2\bigr)C_A C^A\biggr]
+\cdots.
\end{aligned}
\end{equation}
The Higgs potential is independent of $A_\alpha$ and contributes
nothing to this variation. Multiplying the radial eigenvalue
equation by $\rho\psi$ and integrating by parts gives
\begin{equation}
\label{eq:eft11-vector-mass}
2\pi\int_0^\infty\rho\,d\rho\,
\left(\psi'^2+f^2\psi^2\right)
=\omega_b^2\,2\pi\int_0^\infty\rho\,d\rho\,\psi^2
=\omega_b^2.
\end{equation}
The same normalization therefore fixes both kinetic and mass
coefficients, yielding the Proca action
\begin{equation}
\label{eq:eft11-vector-quadratic}
\begin{aligned}
S_V^{(2)}={}&-\frac1{e^2}\int d^2\sigma\sqrt{-\gamma}\\
&\times\left(\frac14G_{AB}G^{AB}
+\frac{\omega_b^2}{2}C_A C^A\right)+\cdots.
\end{aligned}
\end{equation}
No transverse integral or additional normalization factor remains.
The equal scalar and vector masses follow from the BPS relations
in Eq.~\eqref{eq:eft11-profile-equations}.

\subsection{Direct mixed cubic interaction}

The action in Eq.~(\ref{eq:eft-combined-bps-action}) of the main text 
also displays the interaction linear in $\chi$ and quadratic in $C_A$. This 
term comes entirely from
the Higgs kinetic term. Its part quadratic in the vector amplitude
is
\begin{equation}
\label{eq:eft11-higgs-vector-mass}
S\supset-\frac1{2e^2}\int d^2\sigma\,d^2u\,\sqrt{-g}\,
(f+\chi\varphi)^2\psi^2 g^{AB}C_A C_B.
\end{equation}
Expanding the squared Higgs profile and integrating over the normal
plane gives
\begin{equation}
\label{eq:eft11-mixed-cubic}
\begin{aligned}
S_{SV}^{(3)}&=-\frac{g_{SV}}{e^2}
\int d^2\sigma\sqrt{-\gamma}\,\chi C_A C^A+\cdots,\\
g_{SV}&=2\pi\int_0^\infty\rho\,d\rho\,f\varphi\psi^2.
\end{aligned}
\end{equation}
The factor of two in $(f+\chi\phi)^2$ cancels the microscopic
factor $1/2$. The Maxwell action is quadratic in $A_\alpha$, and
the Higgs potential contains no vector field, so neither produces
this direct cubic interaction.

\subsection{Collected action and order of the omitted terms}

Adding the preceding contributions gives
\begin{equation}
\label{eq:eft11-collected}
\begin{aligned}
S_{\rm eff}=-\frac1{e^2}\int d^2\sigma\sqrt{-\gamma}\,
\biggl[&\mu+\nu\mathcal R- \kappa_{\Omega} \Omega_{AB} \Omega^{AB}\\
&+\frac12\nabla_A\chi\nabla^A\chi
+\frac{\omega_b^2}{2}\chi^2\\
&+\frac14G_{AB}G^{AB}+\frac{\omega_b^2}{2}C_A C^A\\
&-\kappa_S\chi\mathcal R
+\frac{\kappa_V}{2}G_{AB}\Omega^{AB}\\
&+g_{SV}\chi C_A C^A\biggr]+\cdots.
\end{aligned}
\end{equation}
This reproduces all displayed terms, signs, and normalizations of
Eq.~\eqref{eq:eft-combined-bps-action} in the main text. The coefficient 
integrals are given in
Eqs.~\eqref{eq:eft11-mu-nu}, \eqref{eq:eft11-shape-tadpole},
\eqref{eq:eft11-vector-tadpole}, and \eqref{eq:eft11-mixed-cubic}.

Finally, the linear profile ansatz also produces a scalar cubic
selfinteraction. The second BPS square has the argument
$\omega_b\psi\chi+\varphi^2\chi^2/2$, so its cubic energy coefficient
is $\pi\omega_b\int_0^\infty\rho\,d\rho\,\psi\varphi^2=g_{SV}/2$.
Hence
\begin{equation}
\label{eq:eft11-scalar-cubic}
S_S^{(3)}\supset-\frac{g_{SV}}{2e^2}
\int d^2\sigma\sqrt{-\gamma}\,\chi^3.
\end{equation}
This term is also included in the displayed main text action.
If the transverse displacement, $\chi$, and $C_A$ are all counted
as first order about a straight string, $\mathcal R$ and
$\Omega_{AB}$ begin at second order. Under that counting,
Eq.~\eqref{eq:eft11-scalar-cubic} must be added to obtain the
complete cubic action within the stated profile ansatz; further 
terms not displayed here begin therefore at fourth order.

\subsection{Relation between $\kappa_V$ and $\kappa_S$}
\label{sub-sec:kappas-relation}

In this subsection, we explain how the two geometric coupling
constants discussed in the previous subsections, $\kappa_V$
and $\kappa_S$, are related. We begin with the definition
of the scalar geometric coupling,
\begin{equation}
\kappa_S=\pi\int_0^\infty d\rho\,
\left(\rho^2f'\varphi+\rho\alpha'a\right).
\end{equation}
Substituting the BPS mode relations from
Eq.~\eqref{eq:eft11-profile-equations} expresses both
shape mode profiles in terms of $\psi$:
\begin{equation}
\kappa_S=\frac{\pi}{\omega_b}
\int_0^\infty d\rho\,
\left(\rho^2ff'\psi-\rho\alpha'\psi'\right).
\end{equation}
We integrate the second term by parts to remove the derivative
from $\psi$. The boundary term $[\rho\alpha'\psi]_0^\infty$
vanishes by regularity at the origin and decay at infinity.
Using the background BPS equations in the form
$(\rho\alpha')'=2\alpha'-\rho^2ff'$, we obtain
\begin{equation}
\label{eq:eft11-scalar-overlap}
\begin{aligned}
\kappa_S
&=\frac{\pi}{\omega_b}\int_0^\infty d\rho\,
\left[\rho^2ff'+(\rho\alpha')'\right]\psi\\
&=\frac{2\pi}{\omega_b}
\int_0^\infty d\rho\,\alpha'\psi.
\end{aligned}
\end{equation}

To relate this expression to $\kappa_V$, let us denote by $H$ the
radial operator appearing in
Eq.~\eqref{eq:eft11-profile-equations}.
The background BPS equations then imply
$H(1-\alpha)=2\alpha'/\rho$.
Multiplying by $\rho\psi$ and integrating by parts twice
transfers the radial derivatives from $1-\alpha$ to $\psi$,
without boundary contributions. The eigenvalue equation
for $\psi$ then gives
\begin{equation}
\label{eq:eft11-magnetic-overlap}
\begin{aligned}
2\int_0^\infty d\rho\,\alpha'\psi
&=\int_0^\infty\rho\,d\rho\,\psi H(1-\alpha)\\
&=\int_0^\infty\rho\,d\rho\,(1-\alpha)H\psi\\
&=\omega_b^2\int_0^\infty\rho\,d\rho\,(1-\alpha)\psi.
\end{aligned}
\end{equation}
Substituting this result into
Eq.~\eqref{eq:eft11-scalar-overlap} and recognizing the
radial integral defining $\kappa_V$, we conclude that
\begin{equation}
\label{eq:eft11-bps-couplings}
\kappa_S=\frac{\omega_b}{2}\kappa_V.
\end{equation}

\section{Effective Action versus Collective Coordinate Approach}
\label{app-CC-vs-EFT}

In Ref.~\cite{Alameda-Calvo:2026dat}, the dynamics of a straight
string was approximated by expanding the microscopic fields in a
finite set of fluctuation profiles with time dependent amplitudes.
Substituting this ansatz into the microscopic action and integrating
over the spatial coordinates yields a reduced mechanical Lagrangian
for these amplitudes, following a sort of collective coordinate approach
widely used in soliton dynamics. Here we establish its connection
to the covariant worldsheet action, incorporating the shape mode
description of Ref.~\cite{Aurrekoetxea:2026xrl} within a unified
framework. 

We demonstrate analytically that the cubic interaction coefficients
agree. The equivalence between the two descriptions is understood
through cubic order, modulo total derivatives and terms proportional
to the free equations of motion. These interactions govern
the leading parametric instabilities and the direct energy exchange
between the two internal modes. This agreement supports interpreting
the worldsheet action as a covariant nonlinear extension of the
collective coordinate description, within the particular set of 
modes that we are considering.

\subsection{Shape mode couplings}

\subsubsection{The shape mode coupling in the EFT approach}

We start by writing the relevant terms in the effective action
\begin{equation}
\label{eq:scc-covariant-action}
S_{\rm eff}
=-\int d^2\sigma \,\sqrt{-\gamma}
\left[
\mu+\frac12\nabla_A\chi\nabla^A\chi
+\frac{\omega_b^2}{2}\chi^2-\kappa_S\chi\mathcal R
\right]+\cdots .
\end{equation}
In particular, the curvature interaction has a positive sign in the
action itself:
\begin{equation}
S_{\chi\mathcal R}
=\kappa_S\int d^2\sigma\,
\sqrt{-\gamma}\,\chi\mathcal R,
\end{equation}

For a small transverse displacement in one direction $X(t,z)$ from a straight
string along the z-direction, the Ricci scalar, to quadratic order is given by
\begin{equation}
\label{eq:etcc-ricci}
R \simeq
(\Box X)^2
-
(\partial_a\partial_b X)(\partial^a\partial^b X),
\end{equation}
where indices are raised with the flat worldsheet metric
$\eta_{ab}=\operatorname{diag}(-,+)$ and
$\Box=-\partial_t^2+\partial_z^2$.

On the other hand, the linearized transverse equation for the 
zero modes of this transverse displacement is
\begin{equation}
\label{eq:etcc-free-transverse}
\Box X=0.
\end{equation}
Using this equation we obtain,
\begin{equation}
\label{eq:etcc-ricci-reduction}
R \simeq
-(\partial_a\partial_b X)(\partial^a\partial^b X)
=
-\frac14\Box^2(X^2).
\end{equation}

Consequently, the leading mixed interaction is
\begin{equation}
\label{eq:etcc-cubic-coupling}
S_{\chi\mathcal R}
\simeq
-\frac{\kappa_S}{4}
\int dt\,dz\,\chi\,\Box^2(X^2).
\end{equation}
Integrating by parts twice and discarding boundary terms yields
\begin{equation}
\label{eq:etcc-integration-by-parts}
S_{\chi\mathcal R}
\simeq
-\frac{\kappa_S}{4}
\int dt\,dz\,\chi\,\Box^2(X^2)
=
-\frac{\kappa_S}{4}
\int dt\,dz\,X^2\Box^2\chi.
\end{equation}

The free bound-state equation,
\begin{equation}
\label{eq:etcc-free-shape}
(\Box-\omega_b^2)\chi=0,
\end{equation}
implies $\Box^2\chi=\omega_b^4\chi$. Thus,
\begin{equation}
\label{eq:etcc-local-vertex}
S_{\chi\mathcal R}
\simeq
-\frac{\kappa_S \omega_b^4}{4}
\int dt\,dz\,\chi X^2.
\end{equation}

To compare to the collective coordinate description, consider
a  periodic interval $0\leq z<L$ and write
\begin{equation}
\label{eq:etcc-amplitudes}
X(t,z)=X(t)\cos(kz),
\qquad
\chi(t,z)=S(t).
\end{equation}
Using
\begin{equation}
\int_0^L dz\,\cos^2(kz)=\frac{L}{2},
\end{equation}
we obtain
\begin{equation}
\label{eq:etcc-reduced-action}
S_{\mathrm{coupling}}
\simeq
-\frac{\kappa_S L \omega_b^4}{8}
\int dt\,S(t)X(t)^2.
\end{equation}
The corresponding mixed cubic term in the reduced Lagrangian is
therefore
\begin{equation}
\label{eq:etcc-reduced-lagrangian}
L_{\mathrm{EFT}}^{(3)}=
-\frac{\kappa_S L\omega_b^4}{8}\,S X^2.
\end{equation}

\subsubsection{The shape mode coupling in the Collective Coordinate approach}

Another approach one can follow is to consider the straight string and
allow for a perturbation of the shape mode with time dependent amplitude
given by $S(t)$ and a zero mode parametrized by $X(t)$. Using the radial 
profiles of these modes already discussed in the main part of the 
text, one can introduce these perturbations into the $4d$ action and
integrate the transverse directions. Following this procedure we 
arrive to an {\it mechanical} Lagrangian of the form,

\begin{equation}
\label{eq:gs-reduced-lagrangian}
\mathcal L_{\rm CC}
\supset \frac{L}{2}(\dot S^2-\omega_b^2S^2)
-\frac{g_s L }{2}SX^2+\cdots .
\end{equation}
so that the relevant coupling term is given by,

\begin{equation}
L_{\mathrm{CC}}^{(3)}=
-\frac{ L g_s}{2}\,S X^2.
\end{equation}
and where the coupling between the shape and the zero mode is obtained by computing 
\begin{equation}
\label{eq:gs-final-integral}
g_s=\pi\int_0^\infty d\rho\,\rho~f'^2
\left[f\varphi +a'+\frac{a}{\rho}\right].
\end{equation}

\subsubsection{Matching the coefficients}

We now show that the collective coordinate coefficient $g_s$
and the geometric coupling $\kappa_S$ satisfy
\begin{equation}
\label{eq:cc-coupling-relation}
g_s=\frac{\omega_b^4}{4}\kappa_S.
\end{equation}
Their radial expressions are
\begin{align}
\label{eq:cc-radial-couplings}
g_s
&=\pi\int_0^\infty d\rho\,\rho(f')^2
\left(f\varphi+a'+\frac{a}{\rho}\right),\\
\kappa_S
&=\pi\int_0^\infty d\rho\,
\left(\rho^2f'\varphi+\rho\alpha'a\right).
\end{align}
We use the normalized profiles of the main text,
\begin{equation}
\label{eq:cc-normalized-profiles}
\varphi=\frac{f\psi}{\omega_b},
\qquad
a=-\frac{\psi'}{\omega_b},
\end{equation}
where $\psi$ satisfies
\begin{equation}
\label{eq:cc-radial-operator}
H\psi=\omega_b^2\psi,
\qquad
H=-\frac{d^2}{d\rho^2}
-\frac{1}{\rho}\frac{d}{d\rho}+f^2.
\end{equation}
Consequently,
\begin{equation}
f\varphi+a'+\frac{a}{\rho}
=\frac{1}{\omega_b}
\left(f^2\psi-\psi''-\frac{\psi'}{\rho}\right)
=\omega_b\psi,
\end{equation}
and the collective coordinate coefficient becomes
\begin{equation}
\label{eq:cc-gs-overlap}
g_s=\pi\omega_b
\int_0^\infty d\rho\,\rho(f')^2\psi.
\end{equation}

To simplify $\kappa_S$, define the background magnetic field
\begin{equation}
\label{eq:cc-background-magnetic-field}
B(\rho)\equiv\frac{\alpha'}{\rho}
=\frac{1-f^2}{2},
\qquad
B'=-ff',
\end{equation}
where the second equality follows from the BPS equations.
Substituting Eq.~\eqref{eq:cc-normalized-profiles} into the
expression for $\kappa_S$ gives
\begin{align}
\label{eq:cc-kappa-overlap}
\kappa_S
&=\frac{\pi}{\omega_b}
\int_0^\infty d\rho\,\rho^2
\left(ff'\psi-B\psi'\right)
\nonumber\\
&=-\frac{\pi}{\omega_b}
\int_0^\infty d\rho\,\rho^2(B\psi)'
\nonumber\\
&=\frac{2\pi}{\omega_b}
\int_0^\infty d\rho\,\rho B\psi.
\end{align}
The boundary term $[\rho^2B\psi]_0^\infty$ vanishes by
regularity at the origin and decay at infinity.

The two remaining overlap integrals are related by the identity
\begin{equation}
\label{eq:cc-magnetic-operator-identity}
HB=2(f')^2.
\end{equation}
To establish it, differentiate the first BPS equation,
$f'=(1-\alpha)f/\rho$, obtaining
\begin{equation}
f''+\frac{f'}{\rho}
=\frac{1-\alpha}{\rho}f'-Bf.
\end{equation}
Together with $B'=-ff'$, this gives
\begin{align}
B''+\frac{B'}{\rho}
&=-(f')^2-f\left(f''+\frac{f'}{\rho}\right)
\nonumber\\
&=-2(f')^2+Bf^2,
\end{align}
which immediately yields
Eq.~\eqref{eq:cc-magnetic-operator-identity}.

Finally, the radial operator $H$ is self-adjoint with respect
to the measure $\rho\,d\rho$. Explicitly,
\begin{equation}
\label{eq:cc-self-adjointness}
\int_0^\infty d\rho\,\rho
\left(\psi HB-BH\psi\right)
=
\left[\rho(B\psi'-\psi B')\right]_0^\infty
=0.
\end{equation}
The boundary term vanishes because $B$ and $\psi$ are regular
at the origin and decay at infinity. Using
$HB=2(f')^2$ and $H\psi=\omega_b^2\psi$, we therefore find
\begin{equation}
\label{eq:cc-overlap-identity}
2\int_0^\infty d\rho\,\rho(f')^2\psi
=
\omega_b^2
\int_0^\infty d\rho\,\rho B\psi.
\end{equation}
Substituting this identity into Eq.~\eqref{eq:cc-gs-overlap}
and comparing with Eq.~\eqref{eq:cc-kappa-overlap} gives
\begin{equation}
\label{eq:cc-final-matching}
g_s
=\frac{\pi\omega_b^3}{2}
\int_0^\infty d\rho\,\rho B\psi
=\frac{\omega_b^4}{4}\kappa_S.
\end{equation}
Thus the two expressions for the mixed cubic coupling agree
analytically at critical coupling, with the mode normalizations
used in the main text. This provides an explicit consistency
check between the effective action and the collective coordinate
description for this interaction, supporting their agreement
on the associated leading parametric instability within their
common regime of validity.

\subsection{Vector mode couplings}

Here we will show the analogous situation for the coupling of the vector mode
to the transverse string displacements.

\subsubsection{The vector mode coupling in the EFT approach}

The relevant terms in the effective action are
\begin{equation}
\label{eq:vcc-covariant-action}
\begin{aligned}
S_{\rm eff}={}&-\int d^2\sigma\,\sqrt{-\gamma}
\\
&\times\biggl[
\mu+\frac14G_{AB}G^{AB}
+\frac{\omega_b^2}{2}C_A C^A
\\
&\qquad
+\frac{\kappa_V}{2}G_{AB}\Omega^{AB}
\biggr]+\cdots .
\end{aligned}
\end{equation}

where
\begin{equation}
\label{eq:vcc-vector-coefficient}
\kappa_V=2\pi\int_0^\infty \rho\,d\rho\,(1-\alpha)\psi.
\end{equation}
For the convention $\omega_A=n_{1\mu}\partial_A n_2^\mu$ of the main
text, the Ricci equation gives, to quadratic order in the static gauge
displacements $X^i(t,z)$,
\begin{equation}
\label{eq:vcc-curvature-expansion}
\Omega_{AB}^{(2)}
=\epsilon_{ij}
(\partial_A\partial_C X^i)(\partial_B\partial^C X^j)
\simeq\frac12\epsilon_{ij}
\Box\!\left(\partial_A X^i\partial_B X^j\right).
\end{equation}
The last step uses $\Box X^i=0$. Here and below, $\simeq$ denotes
equivalence through the mixed cubic order, up to total derivatives and
terms proportional to the free equations of motion.

The free Proca equation is
\begin{equation}
\label{eq:vcc-free-proca}
\partial_A G^{AB}=\omega_b^2 C^B.
\end{equation}
Its divergence implies $\partial_A C^A=0$, and consequently
$\Box G^{AB}=\omega_b^2G^{AB}$. Moving the wave operator in
Eq.~\eqref{eq:vcc-curvature-expansion} onto $G^{AB}$ gives
\begin{align}
S_{G\Omega}^{(3)}
&=-\frac{\kappa_V}{2}
\int dt\,dz\,G^{AB}\Omega_{AB}^{(2)}
\nonumber\\
&\simeq-\frac{\kappa_V\omega_b^2}{4}
\int dt\,dz\,\epsilon_{ij}G^{AB}
\partial_A X^i\partial_B X^j
\nonumber\\
&\simeq\frac{\kappa_V\omega_b^4}{4}
\int dt\,dz\,C^B\epsilon_{ij}X^i\partial_B X^j.
\label{eq:vcc-cubic-representative}
\end{align}
In the last step we integrated by parts once and used
Eq.~\eqref{eq:vcc-free-proca}; the term containing
$G^{AB}\partial_A\partial_B X^j$ vanishes by antisymmetry.

To allow for a general relative spatial phase, choose two orthogonal
transverse standing waves of the form
\begin{equation}
\label{eq:vcc-amplitude-ansatz}
X^1=X(t)\cos(kz),\qquad
X^2=Y(t)\cos(kz+\delta)
\end{equation}
together with
\begin{equation}
C_z = C(t)
\end{equation}
where $\delta$ is constant and $k=2\pi n/L$ with $n\geq1$.
For $0\leq z<L$, their antisymmetric product is

\begin{equation}
\label{eq:vcc-standing-wave-current}
\begin{aligned}
\epsilon_{ij}X^i\partial_z X^j
&= kXY\left[\sin(kz)\cos(kz+\delta)\right.\\
&\qquad\left.-\cos(kz)\sin(kz+\delta)\right]\\
&= -kXY\sin\delta.
\end{aligned}
\end{equation}

Thus the mixed cubic interaction takes the form
\begin{equation}
\label{eq:vcc-eft-mechanical}
L_{\rm EFT}^{(3)}=- Lk
\frac{\omega_b^4\kappa_V}{4}\,CXY\sin\delta.
\end{equation}

\subsubsection{The vector mode coupling in the collective coordinate approach}

For the microscopic reduction~\cite{Alameda-Calvo:2026dat}, use the
gauge-compensated translational profiles and the vector profile:

\begin{equation}
\begin{aligned}
\varphi&=\bar\varphi-q^i\bar D_i\bar\varphi,\\
A_j&=\bar A_j-q^i\bar F_{ij},\\
A_z&=C(t)\psi(\rho),\\
A_t&=0,
\end{aligned}
\end{equation}
together with,
\begin{equation}
q^1=X(t)\cos(kz),\quad q^2=Y(t)\cos(kz+\delta),
\end{equation}
where $i,j=x,y$ and $\bar D_i=\partial_i-i\bar A_i$. The BPS relations
give $\bar D_x\bar\varphi=f'$ and
$\bar D_y\bar\varphi=i f'$, so
\begin{equation}
\label{eq:vcc-higgs-zero-modes}
\delta\varphi=-f'(q^1+iq^2).
\end{equation}
The required cubic term comes entirely from the longitudinal Higgs
kinetic term $-|D_z\varphi|^2/(2e^2)$, where
$D_z=\partial_z-iC\psi$. Its part linear in $C$ and quadratic in
$\delta\varphi$ is
\begin{align}
\label{eq:vcc-microscopic-cubic-density}
\mathcal L_{4d,Cqq}^{(3)}
&=C\psi
\operatorname{Im}\!\left(\delta\varphi^*
\partial_z\delta\varphi\right)
\nonumber\\
&=C\psi(f')^2
\left(q^1\partial_zq^2-q^2\partial_zq^1\right).
\end{align}
The Abelian-Maxwell action is quadratic in the linear field ansatz, and
the Higgs potential contains no vector field; neither supplies an
additional $CXY$ term. Integrating over the transverse plane and the
periodic interval gives
\begin{equation}
\label{eq:vcc-collective-cubic}
L_{\rm CC}^{(3)}=-Lk\,g_v CXY\sin\delta,
\end{equation}
where
\begin{equation}
g_v=2\pi\int_0^\infty\rho\,d\rho\,(f')^2\psi.
\end{equation}

\subsubsection{Matching the coefficients}

Using the already defined radial operator, $H$, 
\begin{equation}
\label{eq:vcc-radial-definitions}
H=-\frac{d^2}{d\rho^2}-\frac1\rho\frac d{d\rho}+f^2,
\qquad H\psi=\omega_b^2\psi,
\end{equation}
and the background magnetic field,
\begin{equation}
B=\frac{\alpha'}\rho=\frac{1-f^2}{2}.
\end{equation}
as well as $f'=(1-\alpha)f/\rho$ and $B'=-ff'$, direct differentiation yields
\begin{equation}
\label{eq:vcc-bps-identities}
H(1-\alpha)=2B,
\qquad HB=2(f')^2.
\end{equation}
The second identity is the same one used in the shape mode calculation.
The operator $H$ is self-adjoint with radial measure $\rho\,d\rho$:
\begin{equation}
\int_0^\infty\rho\,d\rho\,(uHv-vHu)
=\left[\rho(vu'-uv')\right]_0^\infty.
\end{equation}
For $\psi$, $B$, and $1-\alpha$, the relevant boundary terms vanish by
regularity at the origin and decay at infinity. Applying this identity
twice gives
\begin{align}
\int_0^\infty\rho\,d\rho\,(f')^2\psi
&=\frac12\int_0^\infty\rho\,d\rho\,\psi HB
=\frac{\omega_b^2}{2}
\int_0^\infty\rho\,d\rho\,B\psi
\nonumber\\
&=\frac{\omega_b^2}{4}
\int_0^\infty\rho\,d\rho\,\psi H(1-\alpha)
\nonumber\\
&=\frac{\omega_b^4}{4}
\int_0^\infty\rho\,d\rho\,(1-\alpha)\psi.
\label{eq:vcc-overlap-identity}
\end{align}
Multiplying by $2\pi$ establishes
\begin{equation}
g_v=\frac{\omega_b^4}{4}\kappa_V,
\end{equation}
and therefore
\begin{equation}
\label{eq:vcc-final-matching}
L_{\rm CC}^{(3)}=L_{\rm EFT}^{(3)}
=-\frac{Lk\omega_b^4\kappa_V}{4}\,CXY\sin\delta.
\end{equation}

The calculations above establish agreement between the cubic
on-shell interactions obtained from the covariant effective action
and those derived within the collective coordinate approach.
This agreement provides an independent consistency check of the couplings and
demonstrates that, after matching the amplitude definitions, the collective coordinate description
reproduces the corresponding weak field interactions of the effective action through cubic order.

Let us finally note that the correspondence between the EFT and the collective coordinate approach becomes considerably more straightforward for contributions involving exclusively massive internal modes. This is expected, as in both approaches these modes are introduced as fluctuations of the original fields around the background solution, using the same transverse profiles. At leading order in curvature corrections, this leads to the same coupling coefficients and dependence on mode amplitudes. This applies, for instance, to the couplings between the shape and vector modes, as well as to the self-interaction terms of the shape mode.

\section{Nambu--Goto scattering and the induced bound mode response}
\label{app-colliding-wiggles}

We derive the Nambu--Goto solution used to evaluate
the leading geometric sources in the main text.
We first construct a perturbative solution in static
gauge, allowing a direct comparison with the field
theory simulations, and then present an exact solution
in null worldsheet coordinates. From this solution,
we calculate the time delay produced by the nonlinear
interaction of counterpropagating transverse wave
packets, providing an analytic prediction for comparison
with the lattice results. We subsequently 
derive the retarded response of the
internal modes to an arbitrary source and specialize to the Gaussian
profiles used in the numerical experiment. Coordinates and masses are
expressed in the dimensionless units adopted in the main text.

\subsection{Perturbative solution in static gauge}

We choose the physical coordinates $(t,z)$ on the worldsheet and write
the embedding of the string in flat spacetime as
\begin{equation}
\label{eq:ng-static-embedding}
X^\mu(t,z)=\bigl(t,X^1(t,z),X^2(t,z),z\bigr).
\end{equation}

The Nambu--Goto equations of motion in this gauge can be written as
\begin{equation}
\label{eq:ng-static-equation}
\begin{aligned}
(\partial_t^2-\partial_z^2)X^i
={}&-(\partial_zX^j\partial_zX^j)\partial_t^2X^i
\\
&+2(\partial_tX^j\partial_zX^j)\partial_t\partial_zX^i
\\
&-(\partial_tX^j\partial_tX^j)\partial_z^2X^i.
\end{aligned}
\end{equation}
An isolated traveling wave $X^i=F^i(t\pm z)$ solves this equation
exactly. Corresponding solutions also exist in the full nonlinear
field theory~\cite{Vachaspati:1990sk}, providing initial data for the
scattering experiment.

For two counterpropagating disturbances, we define
\begin{equation}
q=t+z,\qquad r=t-z,
\end{equation}
and take the leading solution
\begin{equation}
\label{eq:ng-leading-profile}
X^i_{(1)}=e_+^iF(q)+e_-^iG(r),
\end{equation}
where
\begin{equation}
\label{eq:ng-polarization-conventions}
\begin{aligned}
e_\pm^ie_\pm^i&=1,
\qquad e_+^ie_-^i=\cos\delta,
\\
\epsilon_{ij}e_+^ie_-^j&=\sin\delta.
\end{aligned}
\end{equation}
Thus $F$ propagates toward decreasing $z$ and $G$ toward increasing
$z$. The functions $F$ and $G$ are the physical incoming profiles in
static gauge.

For profiles of amplitude $A$ and characteristic width $L$, we
introduce $\epsilon=A/L\ll1$ and expand
\begin{equation}
\label{eq:ng-amplitude-expansion}
X^i=X^i_{(1)}+X^i_{(3)}+X^i_{(5)}+\cdots,
\qquad X^i_{(n)}=\mathcal O(\epsilon^nL).
\end{equation}
Here $F,G=\mathcal O(\epsilon L)$,
$F',G'=\mathcal O(\epsilon)$, and
$F'',G''=\mathcal O(\epsilon/L)$.
Substituting the leading solution into the nonlinear terms in
Eq.~\eqref{eq:ng-static-equation} gives
\begin{equation}
\label{eq:ng-cubic-equation}
\begin{aligned}
(\partial_t^2-\partial_z^2)X^i_{(3)}
={}&-4e_+^iF''(q)G'^2(r)
\\
&-4e_-^iG''(r)F'^2(q).
\end{aligned}
\end{equation}
The terms depending on $e_+\cdot e_-$ cancel. Define
\begin{equation}
\label{eq:ng-integrated-slopes}
\begin{aligned}
I_+(q)&=\int_{-\infty}^{q}ds\,F'^2(s),
\\
I_-(r)&=\int_{-\infty}^{r}ds\,G'^2(s).
\end{aligned}
\end{equation}
Since $\partial_t^2-\partial_z^2=4\partial_q\partial_r$,
we integrate Eq.\eqref{eq:ng-cubic-equation} with boundary 
conditions that recover the prescribed incoming profiles 
$F$ and $G$ before the collision. The resulting solution, including
the leading nonlinear correction generated by their
interaction, is
\begin{equation}
\label{eq:ng-static-cubic-solution}
\begin{aligned}
X^i(t,z)
={}&e_+^iF(q)+e_-^iG(r)
\\
&-e_+^iF'(q)I_-(r)-e_-^iG'(r)I_+(q)
\\
&+\mathcal O(\epsilon^5L).
\end{aligned}
\end{equation}
For initially separated profiles of compact support, the correction
and its first time derivative vanish before the collision. Gaussian
profiles satisfy this condition asymptotically, with exponentially
small corrections at finite initial separation.

For completeness, the next contribution is
\begin{equation}
\label{eq:ng-static-quintic-solution}
\begin{aligned}
X^i_{(5)}
={}&e_+^i\left[F'G'^2I_++\frac12F''I_-^2\right]
\\
&+e_-^i\left[G'F'^2I_-+\frac12G''I_+^2\right],
\end{aligned}
\end{equation}
where $F$, its derivatives, and $I_+$ are evaluated at $q$, while
$G$, its derivatives, and $I_-$ are evaluated at $r$.

Along the outgoing $F$ disturbance, the $G$ disturbance has passed
and $I_-(r)=I_-(+\infty)$. Equation
\eqref{eq:ng-static-cubic-solution} becomes
\begin{equation}
X^i=e_+^i\left[F(q)-I_-(+\infty)F'(q)\right]
+\mathcal O(\epsilon^5L).
\end{equation}
This is the expansion of a delayed profile
$e_+^iF(t+z-\Delta t_F)$, with
\begin{equation}
\label{eq:ng-asymptotic-delays}
\begin{aligned}
\Delta t_F&=\int_{-\infty}^{\infty}ds\,G'^2(s),
\\
\Delta t_G&=\int_{-\infty}^{\infty}ds\,F'^2(s).
\end{aligned}
\end{equation}
The perturbative calculation determines these delays at leading
order $\mathcal O(\epsilon^2L)$. The exact solution below shows that
Eq.~\eqref{eq:ng-asymptotic-delays} is in fact exact for the specified incoming profiles.

\subsection{Exact solution in null coordinates}

The exact scattering solution can be expressed in null worldsheet
coordinates~\cite{Barbashov:1966frq,Barbashov:1966nvq}. Introduce
coordinates $(u,v)$ and define the embedding implicitly by
\begin{equation}
\label{eq:ng-exact-null-solution}
\begin{aligned}
t+z&=u+I_-(v),
\\
t-z&=v+I_+(u),
\\
X^i&=e_+^iF(u)+e_-^iG(v).
\end{aligned}
\end{equation}
Its tangent vectors are
\begin{equation}
\label{eq:ng-exact-null-tangents}
\begin{aligned}
\partial_uX^\mu
&=\left(\frac{1+F'^2}{2},e_+^iF',\frac{1-F'^2}{2}\right),
\\
\partial_vX^\mu
&=\left(\frac{1+G'^2}{2},e_-^iG',-\frac{1-G'^2}{2}\right).
\end{aligned}
\end{equation}
They satisfy
\begin{equation}
\label{eq:ng-null-constraints}
\begin{aligned}
\partial_uX^\mu\partial_uX_\mu&=0,
\qquad
\partial_vX^\mu\partial_vX_\mu=0,
\\
\partial_u\partial_vX^\mu&=0.
\end{aligned}
\end{equation}
These are the Nambu--Goto equations and constraints in null
coordinates. Our solution given by Eq.~\eqref{eq:ng-exact-null-solution}
satisfies these equations in the small slope regime expressing
a single valued transverse position for the string in the static gauge\footnote{Note that this is not a limitation for the 
solutions and one can find solution beyond this condition \cite{Blanco-Pillado:2025gzs}.
Here we stay away from this regime in order to reduce the bulk radiation from these collisions.}.

Expanding the implicit relations gives
\begin{equation}
\begin{aligned}
u&=q-I_-(r)+G'^2(r)I_+(q)+\mathcal O(\epsilon^6L),
\\
v&=r-I_+(q)+F'^2(q)I_-(r)+\mathcal O(\epsilon^6L).
\end{aligned}
\end{equation}
Substitution into $F(u)$ and $G(v)$ reproduces
Eqs.~\eqref{eq:ng-static-cubic-solution} and
\eqref{eq:ng-static-quintic-solution}.

To determine the delay exactly, follow a feature $u=u_0$ of the
$F$ disturbance. Along this characteristic,
\begin{eqnarray}
&&\frac{dt}{dv}=\frac{1+G'^2}{2},
\qquad
\frac{dz}{dv}=-\frac{1-G'^2}{2},\\
&&\frac{d(t+z)}{dv}=G'^2.
\end{eqnarray}
Its incoming trajectory $t+z=u_0$ becomes
$t+z=u_0+I_-(+\infty)$ after the collision. At a fixed detector,
its arrival is therefore delayed by $I_-(+\infty)$, establishing
Eq.~\eqref{eq:ng-asymptotic-delays}. Interchanging the two profiles
gives the delay of $G$.

The outgoing profiles retain their shapes within the NG approximation. Their delays are
independent of the relative polarization, although the geometry
during the collision depends on $\delta$.

\subsection{Gaussian profiles and time delay}

For Gaussian profiles of the form,
\begin{equation}
\label{eq:ng-gaussian-profiles}
F(s)=G(s)=A\exp\left[-\frac{s^2}{2L^2}\right],
\end{equation}
the quantity $I(s) = \int{ds~F'^2(s)}$ becomes
\begin{equation}
\label{eq:ng-gaussian-integral}
I(s)
=\frac{A^2}{L}\left[
\frac{\sqrt{\pi}}{4}
\left(1+\operatorname{erf}\frac{s}{L}\right)
-\frac{s}{2L}e^{-s^2/L^2}
\right].
\end{equation}
Hence the time delay from Eq.~\eqref{eq:ng-asymptotic-delays} is given by,
\begin{equation}
\label{eq:ng-gaussian-delay}
\Delta t_F=\Delta t_G=\frac{\sqrt{\pi}A^2}{2L}.
\end{equation}

For an isolated traveling wave, the excess Nambu--Goto energy above
that of the straight string is
\begin{equation}
E_F=\frac{\mu}{e^2}\int ds\,F'^2(s),
\end{equation}
The delays can therefore be written as
\begin{equation}
\Delta t_F=\frac{e^2E_G}{\mu}~.
\end{equation}
Restoring physical units gives
\begin{equation}
\label{eq:ng-energy-delay}
\begin{aligned}
\Delta t_F^{\rm phys}
&=\frac{E_G^{\rm phys}}{\mu_{\rm phys}},\\
\mu_{\rm phys}&=\pi\eta^2.
\end{aligned}
\end{equation}
with identical expressions for the $G$ wiggle. This was originally
identify in ~\cite{Barbashov:1966frq,Barbashov:1966nvq}. The same relation between delay and incoming excess energy was obtained for domain wall strings in field theory simulations in Ref.~\cite{Blanco-Pillado:2022rad}.

\subsection{The bound mode response}
 To the order we are considering here the dynamics of the worldsheet
 fields obey, according to our effective theory, the equations,
\begin{equation}
\begin{aligned}
(-\Box_\gamma+\omega_b^2)\chi&=\kappa_S\mathcal R,
\\
(-\Box_\gamma+\omega_b^2)\vartheta
&=-\omega_b\kappa_V\mathcal T,
\end{aligned}
\end{equation}
where
$\Box_\gamma=(-\gamma)^{-1/2}\partial_A
(\sqrt{-\gamma}\gamma^{AB}\partial_B)$.
In the static gauge we get,
\begin{equation}
\begin{aligned}
(\partial_t^2-\partial_z^2+\omega_b^2)\chi
&=\kappa_S\mathcal R+\mathcal O(\epsilon^4),
\\
(\partial_t^2-\partial_z^2+\omega_b^2)\vartheta
&=-\omega_b\kappa_V\mathcal T+\mathcal O(\epsilon^4),
\end{aligned}
\end{equation}
where the curvature sources are evaluated on the leading
transverse solution given by Eq.~\eqref{eq:ng-leading-profile}
, namely,
\begin{equation}
\begin{aligned}
\mathcal R
&=-8F''(t+z)G''(t-z)\cos\delta
+\mathcal O(\epsilon^4/L^2),
\\
\mathcal T
&=4F''(t+z)G''(t-z)\sin\delta
+\mathcal O(\epsilon^4/L^2).
\end{aligned}
\end{equation}
Note that both sources vanish for an isolated traveling wave and
become nonzero when the two disturbances overlap.
Their polarization dependence determines which internal
mode is excited: parallel polarizations source the shape
mode, while orthogonal polarizations source the
pseudoscalar mode.

On the other hand, the two bound modes have the same mass and their
spacetime profile sources during the collision are proportional to 
one another. Therefore, their
responses can be expressed in terms of a single
function $\mathcal U$, which satisfies
\begin{equation}
\label{eq:app-common-source-equation}
\mathcal D\mathcal U=J(t,z),
\qquad
J(t,z)=F''(t+z)G''(t-z),
\end{equation}
where
\begin{equation}
\mathcal D=\partial_t^2-\partial_z^2+\omega_b^2.
\end{equation}
We choose the retarded solution, so that the internal
excitation is generated entirely by the collision.
The coupling strengths and relative polarization then
determine how this common response appears in each mode:
\begin{equation}
\label{eq:app-polarized-bound-response}
\begin{aligned}
\chi&=-8\kappa_S\mathcal U\cos\delta
+\mathcal O(\epsilon^4),
\\
\vartheta&=-4\omega_b\kappa_V\mathcal U\sin\delta
+\mathcal O(\epsilon^4).
\end{aligned}
\end{equation} 

As we showed in Appendix subsection~\ref{sub-sec:kappas-relation}, at critical coupling, $\omega_b\kappa_V=2\kappa_S$, so both
responses have the same coefficient multiplying their respective
polarization factors. Reversing $\delta$ preserves the scalar
response and reverses the pseudoscalar response. The leading
duality relations give
\begin{equation}
C_z=\frac{\partial_t\vartheta}{\omega_b}
+\mathcal O(\epsilon^4),
\qquad
C_t=\frac{\partial_z\vartheta}{\omega_b}
+\mathcal O(\epsilon^4).
\end{equation}

To solve Eq.~\eqref{eq:app-common-source-equation}, define the
spacetime Fourier transform
\begin{equation}
\widetilde J(\omega,k)
=\int dt\,dz\,e^{i\omega t-ikz}J(t,z).
\end{equation}
Then, the retarded solution can be written in terms of this function as,
\begin{equation}
\label{eq:app-retarded-fourier-solution}
\mathcal U(t,z)
=\lim_{\epsilon_R\to0^+}
\int\frac{d\omega}{2\pi}\frac{dk}{2\pi}
\frac{e^{-i\omega t+ikz}\widetilde J(\omega,k)}
{k^2+\omega_b^2-(\omega+i\epsilon_R)^2},
\end{equation}
where $\epsilon_R$ is the retarded regulator. Equivalently, the
spatial Fourier modes satisfy
\begin{equation}
\begin{aligned}
\ddot{\mathcal U}_k+\omega_k^2\mathcal U_k&=J_k(t),
\\
\omega_k&=\sqrt{k^2+\omega_b^2},
\end{aligned}
\end{equation}
with solution
\begin{equation}
\label{eq:app-retarded-mode-solution}
\mathcal U_k(t)
=\frac1{\omega_k}\int_{-\infty}^{t}dt'\,
\sin[\omega_k(t-t')]J_k(t').
\end{equation}
For a source that vanishes after a finite time, extending the upper
limit to $+\infty$ gives the exact outgoing response after that time,
\begin{equation}
\label{eq:app-outgoing-onshell-response}
\begin{aligned}
\mathcal U_{\rm out}(t,z)
=\int\frac{dk}{2\pi}\frac{e^{ikz}}{2i\omega_k}
\bigl[
&e^{i\omega_kt}\widetilde J(-\omega_k,k)
\\
&-e^{-i\omega_kt}\widetilde J(\omega_k,k)
\bigr].
\end{aligned}
\end{equation}
For a source with decaying temporal tails, this expression
approximates the response up to the contribution of the remaining
tail. The outgoing excitation is determined by the source transform
on the massive dispersion relation $\omega^2=k^2+\omega_b^2$.

For the source in Eq.~\eqref{eq:app-common-source-equation}, we obtain
\begin{equation}
\label{eq:app-factorized-source-transform}
\widetilde J(\omega,k)
=\frac12\,
\widetilde{F''}\!\left(\frac{\omega-k}{2}\right)
\widetilde{G''}\!\left(\frac{\omega+k}{2}\right),
\end{equation}
where $\widetilde{F''}(p)=\int ds\,e^{ips}F''(s)$ and similarly
for $G$.

\subsection{Gaussian source and late time response}

For the profiles in Eq.~\eqref{eq:ng-gaussian-profiles},
\begin{equation}
F''(s)
=\frac{A}{L^2}\left(\frac{s^2}{L^2}-1\right)
e^{-s^2/(2L^2)}.
\end{equation}
The common source is therefore
\begin{equation}
\label{eq:app-gaussian-source}
\begin{aligned}
J(t,z)
=\frac{A^2}{L^4}
\left[
\frac{(t^2-z^2)^2}{L^4}
-\frac{2(t^2+z^2)}{L^2}+1
\right]
e^{-(t^2+z^2)/L^2}.
\end{aligned}
\end{equation}
Since
\begin{equation}
\widetilde{F''}(p)
=-\sqrt{2\pi}\,ALp^2e^{-L^2p^2/2},
\end{equation}
Eq.~\eqref{eq:app-factorized-source-transform} gives
\begin{equation}
\label{eq:app-gaussian-source-transform}
\widetilde J(\omega,k)
=\frac{\pi A^2L^2}{16}
(\omega^2-k^2)^2e^{-L^2(\omega^2+k^2)/4}.
\end{equation}
Evaluating this expression on the massive dispersion relation gives
\begin{equation}
\label{eq:app-gaussian-onshell-transform}
\widetilde J(\omega_k,k)
=\frac{\pi A^2L^2\omega_b^4}{16}
e^{-\omega_b^2L^2/4}e^{-L^2k^2/2}.
\end{equation}
Substituting
Eq.~\eqref{eq:app-gaussian-onshell-transform} into
Eq.~\eqref{eq:app-outgoing-onshell-response} yields
\begin{equation}
\label{eq:app-gaussian-outgoing-integral}
\begin{aligned}
\mathcal U_{\rm out}(t,z)
\simeq{}&
\frac{\pi A^2L^2\omega_b^4}{16}
e^{-\omega_b^2L^2/4}
\\
&\times\int\frac{dk}{2\pi}
e^{-L^2k^2/2}\cos(kz)\frac{\sin(\omega_kt)}{\omega_k}.
\end{aligned}
\end{equation}
At this stage, the approximation consists of neglecting the
remaining Gaussian forcing. The integral retains the full massive
dispersion relation and can be evaluated before taking a large time
limit.

At the collision center, $z=0$, the phase has a stationary point at
$k=0$. Expanding
\begin{equation}
\omega_k=\omega_b+\frac{k^2}{2\omega_b}
+\mathcal O(k^4)
\end{equation}
and using
\begin{equation}
\int_{-\infty}^{\infty}\frac{dk}{2\pi}
e^{itk^2/(2\omega_b)}
=\sqrt{\frac{\omega_b}{2\pi t}}\,e^{i\pi/4}
\qquad(t>0)
\end{equation}
gives
\begin{equation}
\label{eq:app-gaussian-late-time-response}
\mathcal U(t,0)
\simeq
\frac{\sqrt{2\pi}}{32}A^2L^2\omega_b^{7/2}
e^{-\omega_b^2L^2/4}
\frac{\sin(\omega_bt+\pi/4)}{\sqrt{t}}.
\end{equation}
This asymptotic form requires
\begin{equation}
t\gg\max\left\{L,\omega_bL^2,\omega_b^{-1}\right\},
\end{equation}
while nonlinear decay and backreaction must remain perturbatively
small over the comparison interval. The factor $t^{-1/2}$ describes
dispersive spreading along the string. The neglected source tail and
the stationary-phase corrections are separate from the amplitude
expansion, whose leading internal response is
$\mathcal O(\epsilon^2)$ at fixed $L$.

\section{Details of the numerical simulations and convergence testing}
\label{app-convergence-test}

In order to solve for the field theory dynamics in flat space,  we implement the equations of motion obtained from Eq. \eqref{eq:FT-action} in \texttt{GRDzhadzha} \cite{Aurrekoetxea:2023fhl}, as described in \cite{Zilhao:2015tya,Helfer:2018qgv}. This is a code by the GRTL Collaboration that incorporates Adaptive Mesh Refinement (AMR) \cite{Radia:2021smk} to accurately resolve the different scales  in the problem: the string core and the length of the string\footnote{See also simulations of global strings with adaptive mesh refinement in Refs. \cite{Drew:2019mzc,Drew:2022iqz,Drew:2023ptp}.}. We extract the values of the scalar and gauge fields and compute the shape and vector modes by projecting these values onto the bound modes of a straight BPS string along the $z-$axis. We follow the procedure of Ref. \cite{Alameda-Calvo:2026dat}, but instead apply it separately on each slice $z=\mathrm{const}$, rather than to the whole volume. We use normalized modes
\begin{equation}
    2\pi\int \rho d\rho\left(\varphi^2 + a^2\right) = 
    2\pi\int \rho d\rho\psi^2 = 1
\end{equation}
with $\rho^2=x^2+y^2$,  so that the expressions for the shape and vector modes are
\begin{equation}
\begin{aligned}\label{eq:chi_projection}
\chi(t,z) &= \int_0^{2\pi} d\theta\int_0^\infty \rho d\rho\left(\delta\vert\phi\vert\varphi + \delta A_\theta a\right)\,,\\
C(t,z) &= \int_0^{2\pi}d\theta\int_0^\infty \rho d\rho A_z\psi
\end{aligned}
\end{equation}
where $\delta\vert\phi\vert\equiv\vert\phi\vert-f(\rho)$ and $\delta A_\theta\equiv A_\theta-\alpha(\rho)/\rho$. In practice, on the AMR grid the discretised string is not exactly the continuum BPS solution we start from, and even with no perturbation, Eq. \eqref{eq:chi_projection} induces some small, grid-dependent $\chi$. For very precise measurements of the shape mode, this noise can be problematic as it propagates from the AMR refinement boundaries. But it can be removed by subtracting, slice by slice and time by time, the $\chi(t,z)$ of a run of the unperturbed string on the same grid. This is not necessary for the vector mode, as the background has $A_z=0$.

\begin{figure}
  \centering
  \includegraphics[width=\columnwidth]{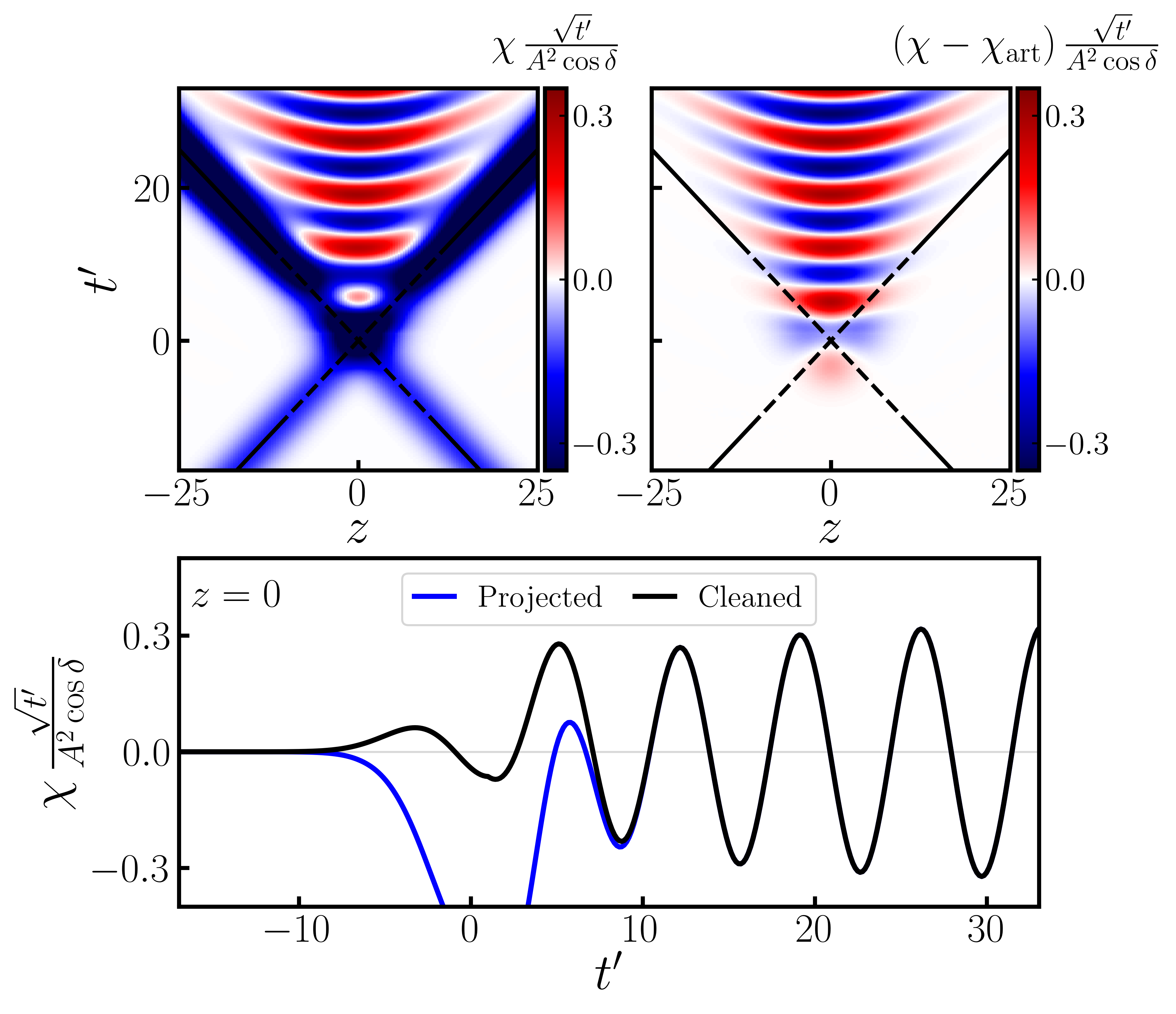}
  \caption{
  Cleaning of the artificial shape mode of the traveling waves. \emph{Top:} the shape mode in spacetime as projected (left), and after subtracting $\chi_\mathrm{art}$ at each slice (right) due to the displacement of the core. \emph{Bottom:} the shape mode at $z = 0$ as projected (blue) and cleaned (black). Time is measured from the collision, $t' = t - t_c$, and the modes are rescaled by $\sqrt{t'}$ for $t'>1$.
  }
  \label{fig:appendix_collision}
\end{figure}

In addition, note that for the initial traveling waves, the projection onto the shape mode leads to an artificial $\chi_\mathrm{art}$ where the waves are located: a displaced core projects onto the shape mode even if it is not deformed. This artificial mode propagates along the string with the traveling waves at the speed of light, as shown in the top left panel of Fig. \ref{fig:appendix_collision}. We can remove it by subtracting $\chi(t,z)\rightarrow \chi(t_,z) - \chi_\mathrm{art}(t,z)$ at each slice $z$. Here $\chi_\mathrm{art}$ is the projection of an
undeformed string whose core is displaced by $\mathbf{d}(t,z)$, so that it is given by Eq. \eqref{eq:chi_projection} with $\vert\phi\vert(t,\mathbf{x}-\mathbf{d})$ and $A_\theta(t,\mathbf{x}-\mathbf{d})$. The displacement of the string $\mathbf{d}(t, z)$ can be measured by finding the zeros of the scalar field $\phi$ at each time $t$ and slice $z$.

We compute numerically the time delay caused by the non-linear interaction of the traveling waves by tracking the position of one of them along the string, and compare it with the Nambu--Goto prediction Eq. \eqref{eq:validation-gaussian-delay}. On each slice $z$, we obtain the transverse displacement of the string $\mathbf{d}=(x_c,y_c)$. At each time, we fit $y_c$ to a Gaussian $\sim \exp[-(z-z_w)^2]$, and take its centre $z_w$ as the position of the wave. The time delay is the lag of $z_w$ behind a light ray leaving the wave's initial position. For the wave traveling towards positive $z$ the time delay is then 
\begin{equation}
    \Delta t(t) = t - \vert z_w(t)- z_w(0)\vert
\end{equation}
The delay vanishes while the wave travels freely, and settles to a constant once the waves have separated, see the bottom panel of Fig. \ref{fig:validation-spacetime-delay}. While the two waves overlap, the fit does not give any meaningful position, so we do not interpret the time delay in this interval.

\begin{figure}[b!]
  \centering
  \includegraphics[width=\columnwidth]{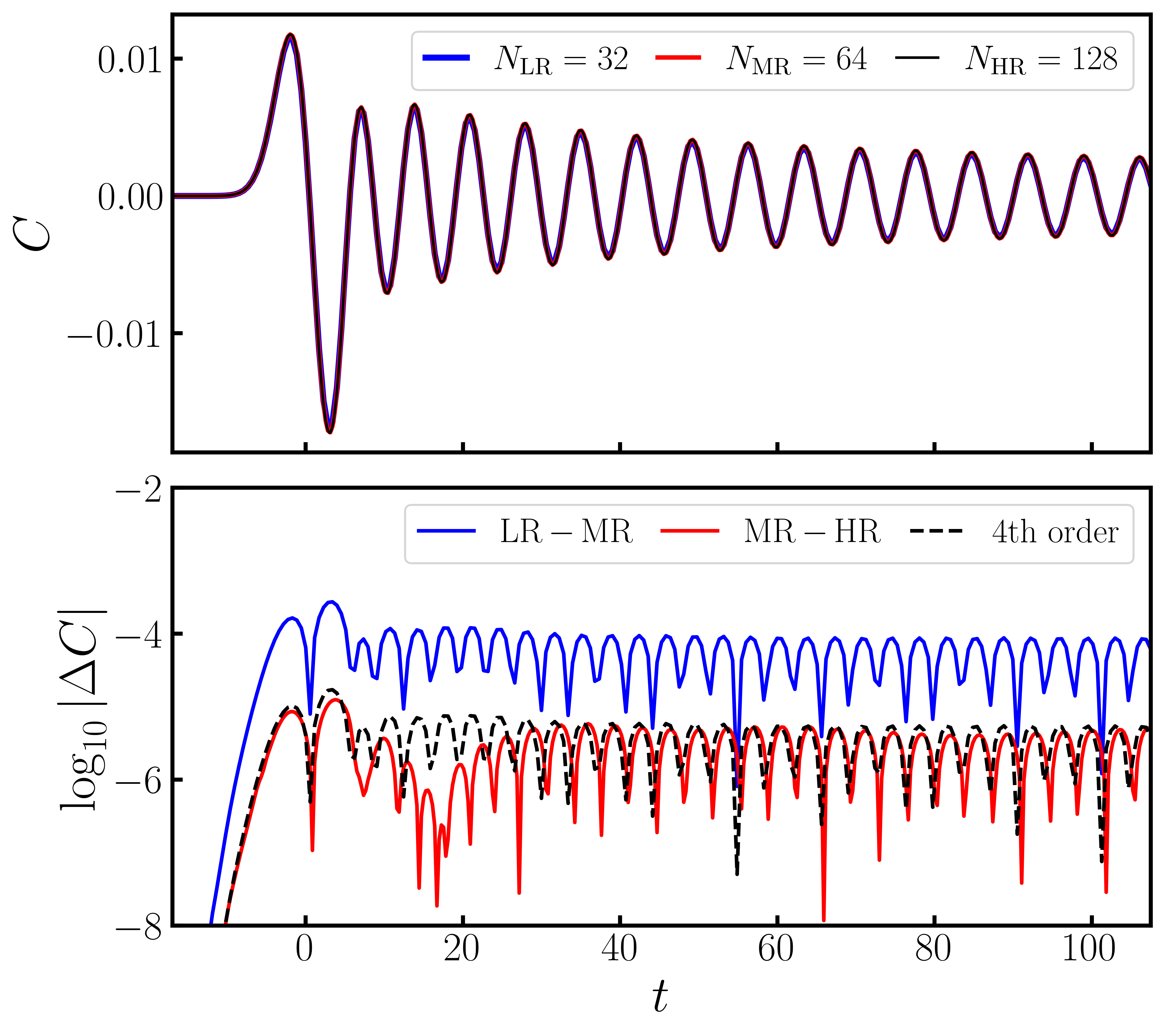}
  \caption{
  Convergence test of the excitation of the vector mode $C(t)$ after the collision of two traveling waves of $A=0.35$, $L=4$ and $\delta=45^\circ$. Low, medium and high resolution correspond to $N=32,\, 64,\, 128$ number of coarse grid points per side. We add three levels of refinement, and our results converge at 4th order.
  }
  \label{fig:appendix_convergence}
\end{figure}

To validate our numerical simulations and results, we perform a convergence test of the excitation of the vector during the collision of traveling waves with ($A,\,L,\,z_w)=(0.35,\,4,\,17)$ and $\delta=45^\circ$. We choose three different resolutions: $(N_\mathrm{LR},\, N_\mathrm{MR},\, N_\mathrm{HR})=(32,\,64,\,128)$ number of points per side in the coarsest level, with three levels of refinement. We use a cubic box of side $128\sqrt{2}$, which leads to the finest grid spacings $(\Delta x_\mathrm{LR}, \Delta x_\mathrm{MR}, \Delta x_\mathrm{HR})\approx (0.71,\, 0.35,\, 0.18)$. In Fig. \ref{fig:appendix_convergence} we compare the errors between resolutions, which show the results converging at 4th order. 

\newpage

\bibliography{Bibliography}

\end{document}